*Dedicated to the memory of Dr. Mikhail I. Eremets*

# Transmission of radio-frequency waves and nuclear magnetic resonance in lanthanum superhydrides

Dmitrii V. Semenok [1,†,*], Florian Bärtl [2,3,†], Di Zhou [1,†], Toni Helm[2,4], Sven Luther[2], J. Wosnitza[2,3], Ivan A. Troyan[5], Viktor V. Struzhkin[1,6,*], and Hannes Kühne[2,*]

[1] *Center for High Pressure Science and Technology Advanced Research (HPSTAR), Beijing*

[2] *Dresden High Magnetic Field Laboratory (HLD-EMFL) and Würzburg-Dresden Cluster of Excellence ct.qmat, Helmholtz-Zentrum Dresden-Rossendorf, 01328 Dresden, Germany*

[3] *Institute for Solid State and Materials Physics, Technische Universität Dresden, 01062 Dresden, Germany*

[4] *Max Planck Institute for Chemical Physics of Solids, 01187 Dresden, Germany*

[5] *A.V. Shubnikov Institute of Crystallography of the Kurchatov Complex of Crystallography and Photonics (KKKiF), 59 Leninsky Prospekt, Moscow 119333, Russia*

[6] *Shanghai Key Laboratory of Material Frontiers Research in Extreme Environments (MFree), Shanghai Advanced Research in Physical Sciences (SHARPS), Pudong, Shanghai 201203, China*

[*]Corresponding authors, emails: dmitrii.semenok@hpstar.ac.cn (Dmitrii V. Semenok), viktor.struzhkin@hpstar.ac.cn (Viktor V. Struzhkin), h.kuehne@hzdr.de (Hannes Kühne).

[†]These authors contributed equally to this work

## Abstract

The discovery of near-room-temperature superconductivity in the lanthanum hydride $LaH_{10}$ has revolutionized this field of research. However, the need to use diamond anvils for the synthesis of such superconductors severely limits the number of experimental techniques to study these materials. Nuclear magnetic resonance (NMR) is one of the key methods for probing spin systems of superconductors. In this work, we show how $^1H$ NMR measurements can be realized in diamond anvil cells to study high-temperature superconductivity in lanthanum polyhydrides at pressures up to 165 GPa. In the newly discovered superhydride $LaH_{12}$, we observed a pronounced suppression of the $^1H$ NMR signal intensity below $T_c^{onset}$ = 260 K in a magnetic field of 7 T, corresponding to the screening of the radio-frequency pulses. Below the critical temperature, all $^1H$ NMR characteristics, including the nuclear spin-lattice relaxation rate $1/T_1T$, demonstrate pronounced features, evidencing the bulk nature of the superconducting transition. In zero field, the radio-frequency signal transmission through the $LaH_{12}$ sample shows a pronounced drop below $T_c^{onset} \approx 267$ K, further confirming superconductivity in the sample. A description of the $1/T_1T$ data with an exponential form allows the estimation of the superconducting gap Δ(0) lying between 427 and 671 K (corresponding to 36.8 to 57.8 meV), and the ratio $R_\Delta = 2\Delta(0)/k_BT_c$ between 3.76 and 5.16 in the synthesized hydride sample.

## Introduction

Superhydrides are a rapidly developing class of hydrogen-rich compounds that have attracted considerable attention in the field of condensed-matter physics due to their remarkable superconducting properties. The discovery of high-temperature superconductivity (SC) in $H_3S$ ($T_c^{max}$ = 203 K[1]), $LaH_{10}$ ($T_c^{max}$ = 250 K[2,3]), $ThH_{10}$ ($T_c^{max}$ = 161 K[4]), $YH_6$ ($T_c^{max}$ = 224-226 K[5,6]), $CeH_{10}$ ($T_c^{max}$ = 115 K[7]), and many other polyhydrides[8,9] has opened new avenues for fundamental and applied studies. Particularly, the search for room-temperature superconductors and the development of magnetic field sensors, diodes, and memory elements based on superhydrides gave a new impetus to the research efforts[10-12]. Unfortunately, polyhydride samples are often inhomogeneous[13,14]. A comprehensive study of SC in such inhomogeneous samples is problematic due to the presence of several crystallographic phases, including metastable ones [15], with very different critical temperatures (Fig. 1a). Indeed, impurity phases or grains, separated from the probing current trajectories by non-superconducting lower hydrides or insulating shells of higher polyhydrides, cannot be identified by standard four-probe transport measurements (Fig. 1a). Such grains with varying superconducting properties may be responsible for broadened or steplike transitions in the temperature dependence of the resistance, hardly distinguishable inhomogeneous sample quality, or stress distributions. Detection methods based on volume-penetrating electromagnetic fields can be suitable probing techniques in this situation. Radio-frequency (RF) contactless measurements[16,17] and nuclear magnetic resonance (NMR) spectroscopy[18] allow to overcome this problem and detect SC in inhomogeneous samples under pressure.

NMR spectroscopy is based on the coherent manipulation of the nuclear spin system in a material[19,20]. It can be employed to probe the magnetic hyperfine fields, originating from the electronic subsystem. Thus, it is very sensitive to the static electronic spin susceptibility and low-energy excitations of the electronic system. The shape of an NMR spectrum, its broadening and frequency shift depend on the magnetic interactions of the nuclear moments with their environment[21]. $^{27}$Al-NMR spectroscopy played a major role in confirming the Bardeen–Cooper–Schrieffer (BCS) theory of SC[22]. Moreover, Redfield, Anderson, Hebel, and Slichter discovered that at temperatures below the superconducting critical temperature ($T_c$), there is a maximum of the $^{27}$Al spin-lattice relaxation rate $1/T_1$ (the Hebel-Slichter peak) in aluminum[23-25]. Measurements of $1/T_1$ at temperatures well below $T_c$ yield an exponential decrease of $1/T_1$ with temperature in the case of a full superconducting gap. In metals, the NMR frequency shift (also called Knight shift[26]) and the spin-spin relaxation rate $1/T_2$ are independent of temperature[27], whereas $1/T_1$ is proportional to temperature (known as Korringa relation[28]). A distinctive feature of a superconducting transition is a sharp decrease of the relaxation rate $1/T_1T$ due to the opening of the superconducting gap.

Since the beginning of research on superhydrides, it was suggested that they were the most prominent representatives of BCS superconductors[1]. NMR spectroscopy is a powerful tool for studying the nature of the normal and superconducting ground states. Therefore, it can help to obtain information on the superconducting gap in the polyhydrides. However, NMR studies on these materials face a major experimental challenge—the tiny sample space of a diamond anvil cell (DAC), which is required to create the necessary pressures of over 1 million atmospheres (100 GPa). Here, the comparably high sensitivity of NMR spectroscopy to the $^1$H isotope, the sublattice of which plays a major role in the SC of hydrides[8], is very beneficial. Fortunately, an approach has recently been developed that allows the detection of ultra-small samples via conventional solid-state NMR techniques, using focusing Lenz lenses deposited on diamond anvils[18,29,30]. Lower hydrides of FeH[31,32], $Cu_2H$[33], $YH_3$ [18], $LaH_3$ [30], and molecular hydrogen[34,35], have previously been studied at pressures up to 220 GPa. Below, we will demonstrate that the same Lenz lenses also prove to be a very efficient tool for studying the surface impedance of superhydride samples. Unlike NMR, within this RF transmission method, two Lenz lenses are used separately as excitation (A) and detection (B) microcoils of an RF transformer (Fig. 1a).

In this work, we applied RF transmission and $^1$H NMR techniques to study lanthanum superhydrides, synthesized from La or $LaH_{\sim 3}$ and ammonia borane ($NH_3BH_3$) in five DACs (labeled as "N0-N4") at different pressures up to 165 GPa. We were able to study the $^1$H NMR signal of various lanthanum polyhydrides and observed the superconducting transition stemming from the newly discovered (pseudo)hexagonal $LaH_{12}$, yielding a $T_c^{onset}$ of at least 267 K. This is confirmed by an abrupt change in the RF surface impedance and a sharp drop of the $^{1}$H NMR spin-lattice relaxation rate.

## Results and discussion

### *1. Combined DAC with electrodes and Lenz lens*

Although the RF method of measuring the surface impedance has been known since 1989 [16], and has been widely used to probe superconducting transitions[17,36], it has not been applied in combination with Lenz lenses[37] for high-pressure studies. To bridge this gap, in the first (calibration) experiment utilizing the DAC N0 (Fig. 1a-e), we used two detection systems simultaneously: a classical four-probe van der Pauw resistance measurement circuit was deposited on one diamond anvil, while a simple single-stage Ta/Au Lenz lens (Fig. 1b) was sputtered onto the opposite anvil. After laser heating of La microparticles loaded with ammonia borane ($NH_3BH_3$) at 149 GPa, the pressure in the DAC dropped to 120 GPa, and the four-contact measurements showed a pronounced resistance drop starting below $T_c^{onset}$ = 205 K (Fig. 1d). This $T_c$ is in good agreement with previous studies of this La superhydride[38]. The result of our synthesis may correspond to the polyhydride $LaH_{9+x}$ ($x$ = −1…+1) with a low symmetry (e.g., $C2/m$[38]) and, partially, a molecular hydrogen sublattice[39]. A detailed description of the preparation and characteristics of the sample in DAC N0 is given in the Supporting Table S1.

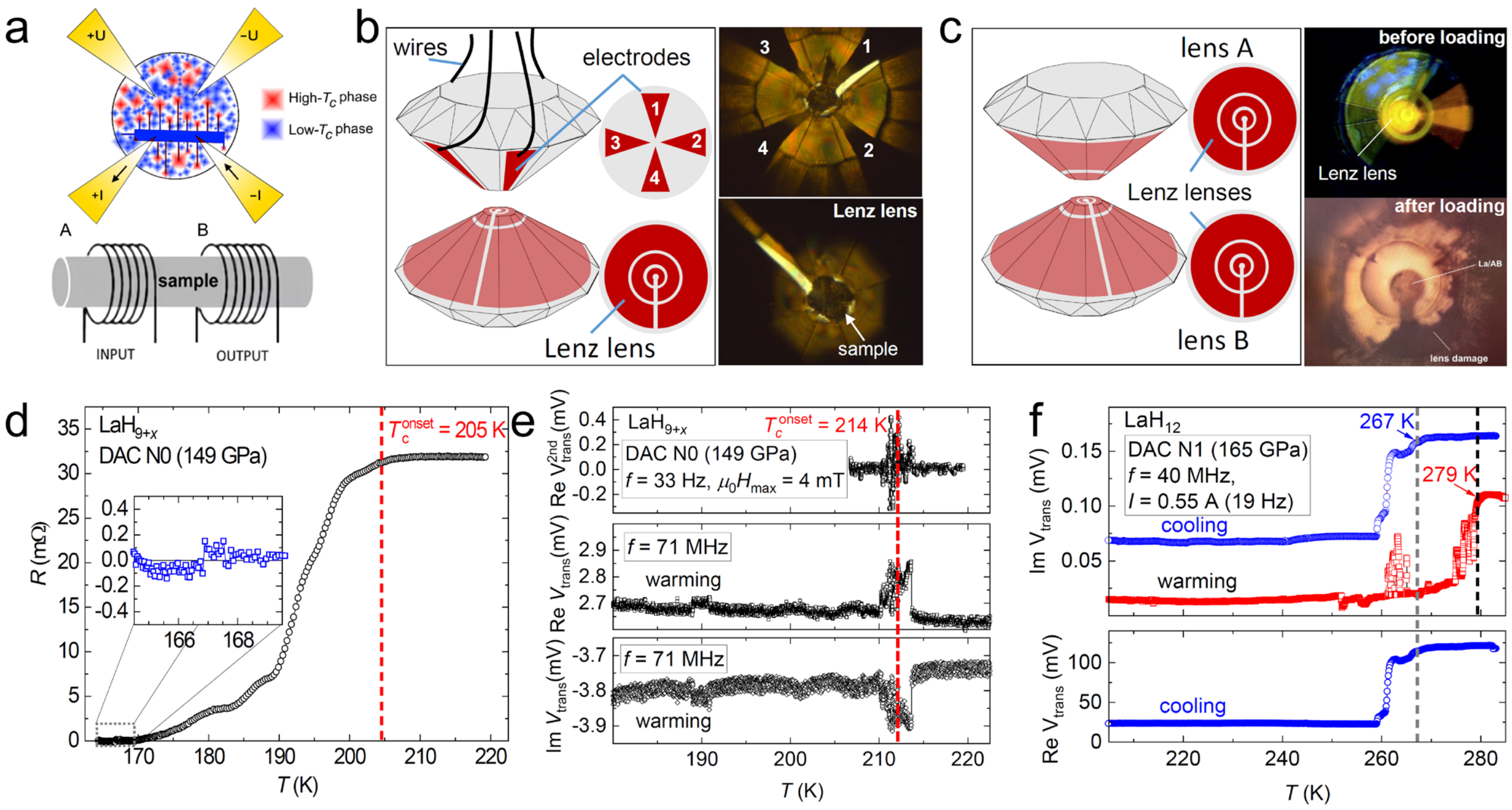

**Figure 1. Electrical resistance and RF transmission of lanthanum superhydrides, synthesized from La and $NH_3BH_3$ in the DACs N0 and N1.** (a) Sketch of a possible distribution of high-$T_c$ and low-$T_c$ superconducting phases in a hydride sample, which complicates the detection of high-$T_c$ SC using a four-electrode van der Pauw circuit. At the same time, only the low-$T_c$ transition may be detected due to the presence of a homogeneous segment of the corresponding phase (marked in blue) between the electrodes. Also shown is a scheme of RF diagnostics, employing a transformer circuit (coil A is input, coil B is output). (b) Schematic diagram of the DAC N0, containing a four-electrode circuit for transport measurements and a Lenz lens for RF transmission measurements. Right: Images of the DAC culet, loaded sample, electrode system and Lenz lens. (c) Schematic diagram of the DAC N1, containing two Lenz lenses for RF transmission measurements and NMR studies. Right: Images of the anvil culet with sputtered Lenz lenses before and after loading and compression to 165 GPa. (d) Electrical resistance (current is 1 mA, DC) as a function of temperature for a sample of lanthanum hydride in DAC N0 at 120 GPa. Inset: residual resistance of the sample below 170 K. (e) Real (Re) and imaginary (Im) components of the RF signal passing through the DAC N0 sample in transformer mode. The carrier frequency is 71 MHz. Top panel: Directly demodulated RF signal of the 2$^{nd}$ harmonic of the low-frequency modulating field (33 Hz, $B_{max}$ ≈ 4 mT). (f) High-frequency (40 MHz) study of the DAC N1 at 165 GPa. Real and imaginary components of the receiving coil signal in the cooling cycle. The superconducting transition of the sample in DAC N1 starts around 267 K. A thermal hysteresis (about 10 to 12 K) in the system as exemplified by the imaginary (Im) components of the receiving coil signal. The thermometer was glued to the outside of DAC N1.

Using any electrode pair of the DAC N0 as a RF antenna and the Lenz lens with a surrounding, macroscopic single-turn coil as a receiver (Supplementary Figs. S1, S2) allowed us to detect a sharp change in the surface impedance of the sample in the real and imaginary components of the RF signal. This corresponds to a SC transition with an onset at 212 to 214 K (Figure 1e). We detected such steps or jumps of the transmitted RF signal using various frequencies between 0.2 and 71 MHz (Fig. 1e, Supplementary Fig. S3), confirming the reproducibility. For additional verification, we used the double-modulation technique, similar to that proposed by Timofeev in 2002 [17]. As shown in the top panel of Fig. 1e, placing the sample in a weak ($B_{max} \approx 4$ mT) oscillating magnetic field with a frequency of 33 Hz leads to a pronounced feature in the 2$^{nd}$ harmonic (66 Hz) of the directly demodulated RF signal. This feature emerges simultaneously with the jump in the transmitted RF signal, because the magnetic field suppresses SC, regardless of the sign of the magnetic induction $B_z$.

The $T_c$ observed by the RF method is somewhat higher than $T_c^{onset}$, detected my means of the resistance measurements. This is because the high-frequency field penetrates the entire volume occupied by the $LaH_{9+x}$ sample, and not only the space between electrodes, as in the case of the four-electrode circuit. That is why it is possible to detect superconducting phases with the highest $T_c$ regardless of how far they are from the electrodes. Note that reducing the carrier frequency to 0.2-1 MHz also allows for reliable detection of the signal of only the "main" superconducting phases, located near the electrodes, with $T_c$ at about 180 to 190 K (Supplementary Fig. S3).

*2. RF detection of superconductivity with two Lenz lenses*

After successful testing of the contactless RF detection method using Lenz lenses (DAC N0), we prepared DACs containing two Lenz lenses as indicated in Fig. 1c (see supplementary information) for combined RF transmission and $^1$H NMR studies. We performed experiments with four high-pressure DACs (pressure is given in brackets): DAC N1 (165 GPa), N2 (147 GPa), N3 (87 GPa), and N4 (19 GPa). The samples in the DACs N1 and N2 were obtained by infrared laser heating of metallic La in an $NH_3BH_3$ environment, while the samples N3 and N4 were obtained by heating of $LaH_{3-x}$ with $NH_3BH_3$ under pressure. We performed the laser heating at the ID27 beamline of the European synchrotron radiation facility (ESRF). The most interesting results were obtained for the samples N1 and N3, synthesized in symmetric BX-90 mini cells[40], equipped with diamond anvils with a culet diameter of 75 µm. 1-2 µm thick copper or silver films, sputtered onto the diamond anvils, were used for the Lenz lenses[37] (Fig. 1c, Supplementary Fig. S1). As we will demonstrate in the next paragraph, the sample in DAC N1 is best described by the formula $LaH_{12}$. Detailed information on the DAC designs and preparation is given in the Supporting Table S1.

First, we performed RF transmission measurements of the sample in DAC N1 at 165 GPa, employing a high-frequency transformer formed by the two Lenz lenses. As can be seen in Fig. 1f (as well as in Supplementary Fig. S4), the RF signal transmission exhibits main features at 267 to 286 K, which appear in both the amplitude and phase of the output signal of the RF transformer for various frequencies (40, 150, and 200 MHz). This behavior is accompanied by the appearance of a second harmonic at the frequency of the external modulation field (Supplementary Fig. S4). Another, small feature in the RF transmission occurs as well at 248 to 252 K, which probably corresponds to a $LaH_{10}$ impurity (Supplementary Fig. S5). Previous works on lanthanum[3,41-43] and lanthanum-scandium[12] hydrides indeed evidenced an additional step-like drop in the electrical resistance, setting in at temperatures between 265 and 280 K or even above. However, such a high $T_c$ could not be attributed to any known La hydride until now. In the present work, we demonstrate that this high-$T_c$ hydride phase can be detected by contactless methods, in particular by changes in the surface RF impedance.

The RF method reported here, indicates that SC in the La-H system under pressure may occur already at temperatures as high as 267 to 286 K[12,42].

*3. X-ray diffraction of the synthesized lanthanum superhydrides*

Diffraction of synchrotron radiation is an effective technique for elucidating the structure of polyhydrides under high pressure. The structures of the lanthanum superhydrides in the DACs N1, N2, and N3 were studied using powder X-ray diffraction (XRD), presented in Fig. 2). In the case of DAC N1, our XRD analysis shows the formation of a new crystal modification of $LaH_{12}$ at 165 GPa, most likely with a hexagonal *P*6 symmetry (Fig. 2a). In general, an interpretation of the XRD data of the sample in DAC N1 at 165 GPa is challenging. The resulting pattern differs from all the previously experimentally studied ones for the La-H system, but has many similarities with the recently discovered $(La,Ce)H_{12}$ [10]. In terms of volume, the obtained compound approaches the one of $LaH_{12}$ [2].

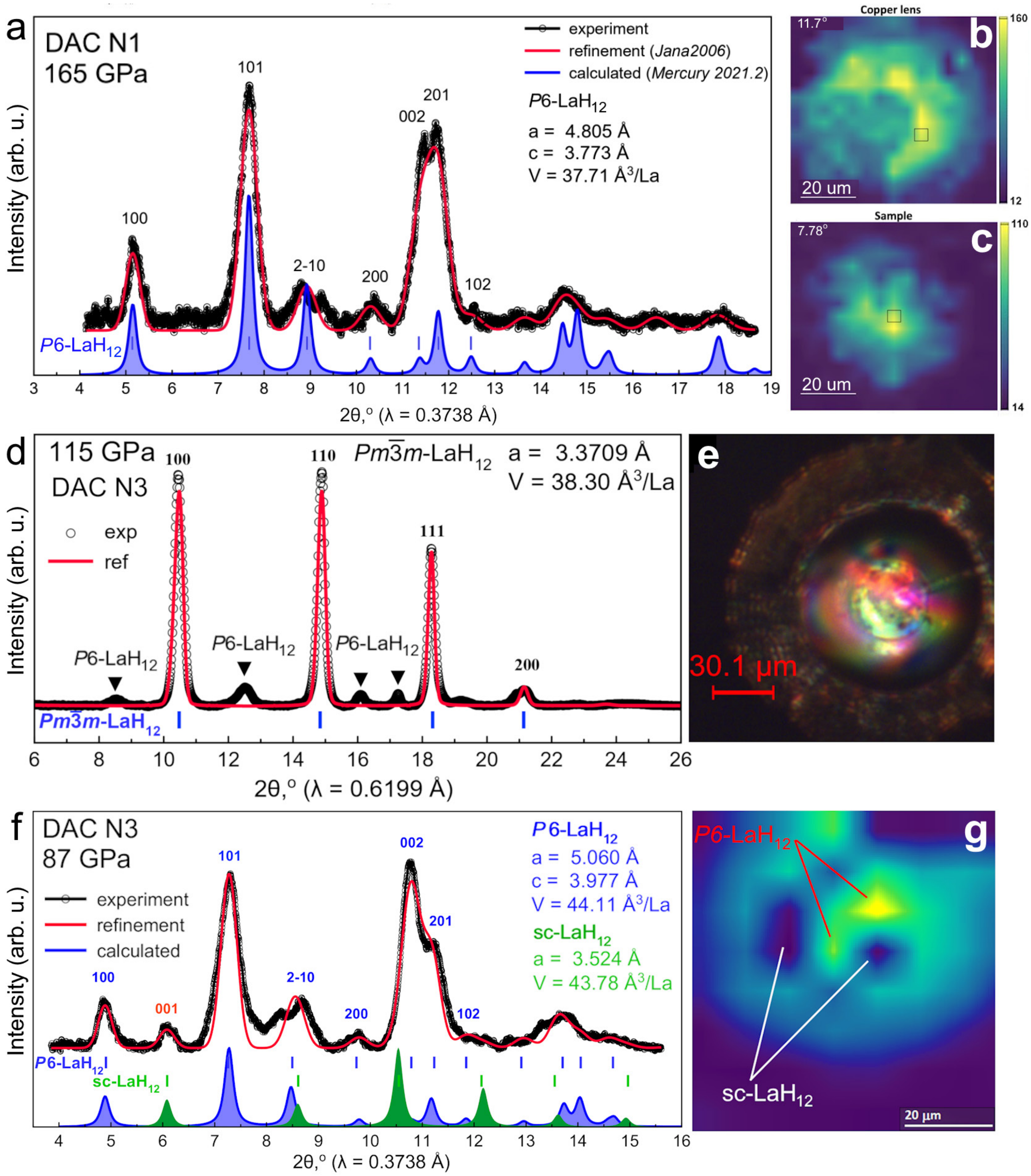

**Figure 2. X-ray diffraction studies of the samples in DAC N1 at 165 GPa and in DAC N3 at 87-115 GPa.** (a) X-ray diffraction pattern of the sample in DAC N1 and Le Bail refinement of the $P6$-$LaH_{12}$ unit-cell parameters. Black circles are the experimental data, the red line is the Le Bail refinement, and the blue line is the XRD pattern calculated using the Mercury 2021.2 software[44]. For the refinement, we used the $P6_3/mmc$-$LaH_9$ structure, which reproduces almost all diffraction peaks, but has an unusually small *c/a* ratio of 0.78. The unit-cell volume corresponds to a hydrogen content of about 12 H atoms per La atom. (b) X-ray diffraction mapping of a copper Lenz lens, and (c) the $P6$-$LaH_{12}$ distribution, obtained using the XDI software[45]. The color scale corresponds to the intensity of the strongest XRD reflection for each compound. (d) X-ray diffraction pattern of the sample in DAC N3 and Le Bail refinement of the $Pm\overline{3}m$-$LaH_{12}$ unit-cell parameters. The sample also contains an impurity fraction of hexagonal $P6$-$LaH_{12}$. Black circles are the experimental data, the red line is the Le Bail refinement. (e) Picture of the DAC N3 culet with the sample. (f) XRD pattern of the sample in DAC N3 after decompression to 87 GPa. Black circles are the experimental data, the red line is the Le Bail refinement, and the blue and green filled curves are the XRD patterns of both $LaH_{12}$ modifications, calculated using the Mercury 2021.2 software[44]. (g) X-ray diffraction map of the $P6$- (yellow-green) and $Pm\overline{3}m$-$LaH_{12}$ (dark blue) spatial distributions (XDI software[45]).

Although the $P6_3/mmc$ structure, with the lattice parameters $a = 4.805$ Å and $c = 3.773$ Å, and a unit-cell volume of $V = 37.71$ Å$^3$/La, describes the observed diffraction pattern quite well (Fig. 2a-c), it should be noted that a *c/a* ratio of 0.78 is unusually small for the $P6_3/mmc$ space group. Moreover, the reflections 002 and 201 have unexpectedly high intensities. Hence, another (lower) space group may also describe the structure of the obtained compound. We set the amount of hydrogen in $P6$-$LaH_{12}$ by analogy with the previously reported simple cubic (sc) $Pm\overline{3}m$-$LaH_{12}$ [2,46-48] with an accuracy of ± 1 H atom.

The sample in DAC N3, consisting of two phases, was initially synthesized at a pressure of about 115 GPa from $LaH_{3-x}$ and $NH_3BH_3$. The first phase detected after the pressure synthesis (115 GPa) was simple cubic ($Pm\overline{3}m$) sc-$LaH_{12}$ (Fig. 2d-e), as reported previously[2,46-48]. The second phase, distributed in close proximity to $LaH_{12}$, is the hexagonal modification $P6$-$LaH_{12}$, which is the main product of the synthesis in DAC N1. This $P6$ phase has almost the same volume as the cubic one (Fig. 2f-g). Both synthesized modifications of $LaH_{12}$ generally retain the structure of their La sublattices upon decompression to 87 GPa (Fig. 2f).

DAC N2, made from BeCu alloy, was loaded with metallic La and $NH_3BH_3$, and heated via a series of IR laser pulses at 147 GPa at the ID27 station of the ESRF. After the laser heating, the pressure increased to 150 GPa. For this sample, we used diamond anvils with a culet diameter of 100 µm. As will be discussed below, the $^1$H NMR signal of the sample in DAC N2 is several times smaller compared to the spectral intensity observed for the samples in in DACs N1 and N3. This likely is caused by a partial destruction of one of the Lenz lenses, as well as a smaller amount of synthesized superhydrides.

According to the XRD analysis (Supplementary Fig. S13), there are two small regions of ~10 µm inside of DAC N2, containing mainly $I4/mmm$-$LaH_4$ and $Cmcm$-$LaH_3$. The first of these compounds is a superconductor with a $T_c$ of about 80 to 90 K[49,50], while the second one is an analogue of the previously found $Cmcm$-$LaH_{\sim 4}$ [14] with a 2 Å$^3$/La smaller unit-cell volume. This corresponds to a $LaH_3$ composition. The $T_c$ of tetragonal $LaH_4$ was reported to decrease rapidly in applied magnetic field ($dT_c/dB \approx -2$ K/T[49]). Therefore, $T_c$ should be 66 to 76 K at 7 T. Due to the low content of superconducting phase and the large spurious background signal with features below 100 K, we cannot confidently state that we have detected the anticipated superconducting transition in $LaH_4$. However, the disappearance of a series of $^1$H NMR signals, and a step in the $1/T_1$ vs $T$ dependence around 60 to 70 K may indicate the emergence of SC in $LaH_4$ (Supplementary Fig. S21).

DAC N4 was loaded with $LaH_{3-x}$ and $NH_3BH_3$. The material was heated by an IR laser at around 150 GPa, but it broke down in the following days with a pressure drop to below 50 GPa. Due to this, DAC N4 was used only for a narrow set of experiments (Supplementary Fig. S22).

*4. Preliminary NMR experiments*

$^1$H NMR experiments on micron-sized samples compressed between diamond anvils are inevitably associated with parasitic hydrogen signals from different parts of the high-pressure DACs, such as the organic adhesive used to glue the diamond anvils to the seats, wire insulation, capacitors, soldering paste, and the printed circuit board (PCB) supporting the excitation coils. Since the laser heating of the starting material in the DACs never leads to a complete decomposition of $NH_3BH_3$, the study of the residual signal of this compound is important as well. Therefore, before starting the experiments with the polyhydrides in the DACs N1-N4, we investigated several calibration samples at temperatures from 10 to 300 K. We studied the following test samples:

(A) An empty DAC with a Lenz-lens system (Supplementary Fig. S8);

(B) The same DAC, loaded with a piece of $^{27}$Al foil and $NH_3BH_3$ (Supplementary Figs. S6-S7);

(C) A bulk sample of $NH_3BH_3$ in a conventional NMR resonator (Supplementary Fig. S9);

(D) A piece of PCB, used as support for the Lenz-lens excitation coils (Supplementary Fig. S10).

The examination of the empty DAC with Lenz lenses (test "A") indicates the presence of some volatile H-containing impurities in the DAC: The NMR signal intensity significantly decreases with time. The low intensity of the $^1$H NMR signal of the empty DAC prevents a $T_1$ measurement (Supplementary Fig. S8).

As a result of the preparatory experiment "B", we found that the Lenz-lens system does indeed allow a reliable detection of the $^{27}$Al NMR signal from a small piece of Al foil from the DAC sample volume. The spin-lattice relaxation curve $I(\tau)$ and the stretching exponent indicate a non-uniform distribution of the RF field in the sample (Supplementary Fig. S6), which in this case is due to a non-ideal geometry of the Lenz lenses in the non-pressurized DAC.

The $^1$H signal intensity was found to sufficiently strong when we placed a micron-sized sample of $NH_3BH_3$ together with the $^{27}$Al foil in the chamber of the test DAC "B" (Supplementary Fig. S6, S7). Besides the presence of a kink in the temperature dependence of the integral intensity $I(T)$ around 250 K, the nuclear spin-lattice relaxation rate $1/T_1T$ yields no particular features between 50 and 300 K, which is important to note for the subsequent experiments on the lanthanum hydrides.

The $^1$H NMR experiments on a bulk sample of $NH_3BH_3$ (test "C") yield a rather complex, two-component $T_1$ relaxation. For instance, $1/T_1T$ shows a rise below 210 K, with a maximum at 100 to 150 K (Supplementary Fig. S9). This is possibly related to low-temperature phase transitions in $NH_3BH_3$, as reported previously (see also Ref.[51]).

Finally, the teflon PCB (test "D") gives an insignificant $^1$H NMR signal without pronounced features above 170 K (Supplementary Fig. S10). Thus, by performing these preliminary experiments, we have verified that the $^1$H NMR signal discussed below indeed dominantly stems from the samples of compressed lanthanum polyhydrides.

*4. $^1$H NMR of lanthanum superhydrides at pressures up to 165 GPa*

We used a system of Lenz lenses (Fig. 3a-b), deposited on diamond anvils, to perform NMR measurements of micron-sized samples under pressure. Our $^1$H NMR study, using DAC N1 in a magnetic field of 7 T ($\nu_0$ = 298 MHz), employing a conventional spin-echo pulse sequence, was carried out during cooling between 300 and 10 K (Fig. 3c). At high temperatures, the spectrum can be described as a composition of at least four peaks ("a-d", Fig. 3d), three of which ("a-c") disappear below 230 K and yield a negative frequency shift with decreasing temperature ($d\nu/dT > 0$). These peaks are attributed to different lanthanum superhydrides that form at 165 GPa, when the La/ $NH_3BH_3$ mixture is heated[2,14,39]. As the XRD analysis demonstrated (Fig. 2), the main phase synthesized in DAC N1 is hexagonal $P6$-$LaH_{12}$. The very narrow peak "d" disappears below 200 K with $d\nu/dT < 0$.

Between 300 K and $T_c^{onset}$ = 260 K, we find a linear decrease of the integral intensity of the $^1$H NMR spectrum, as well as a pronounced slope change at $T_c^{onset}$ (Fig. 3e). Below $T_c^{onset}$, our data show a strong suppression of the NMR intensity, which saturates at 211 K. Between $T_c^{onset}$ and 220 K, we observe pronounced changes in all spectral characteristics, i.e., the frequency shift, spectral width, and area ("intensity") of the $^1H$ NMR spectrum (Fig. 3d-f). The residual, broad low-temperature $^1H$ NMR spectrum stems from the background of non-superconducting hydrogen-containing compounds, residual $NH_3BH_3$, products of its thermal decomposition, and lower La hydrides (Fig. 3c,d). We note that, although these background signals may partially contribute to the change in spectral characteristics in the superconducting transition region, this does not affect the qualitative behavior in the detection of the superconducting transition by means of NMR. At 165 GPa and 7 T, the observed $T_c^{50\%}$ is (227 ± 5) K (Fig. 3f). This temperature corresponds to the SC transition in the best-known lanthanum superhydrides in a field of 7 T, determined via electrical-transport measurements at the middle of the resistance drop ($R_{50\%}$ criterion, see Ref.[2]). At 4 T, the superconducting transition of the lanthanum superhydride in DAC N1 is shifted higher in temperature, to $T_c^{50\%}$ = (235 ± 5) K (Fig. 3g). The offset $T_c$ also increases from 211 K (7 T) to 220 K (4 T), see Supplementary Fig. S18.

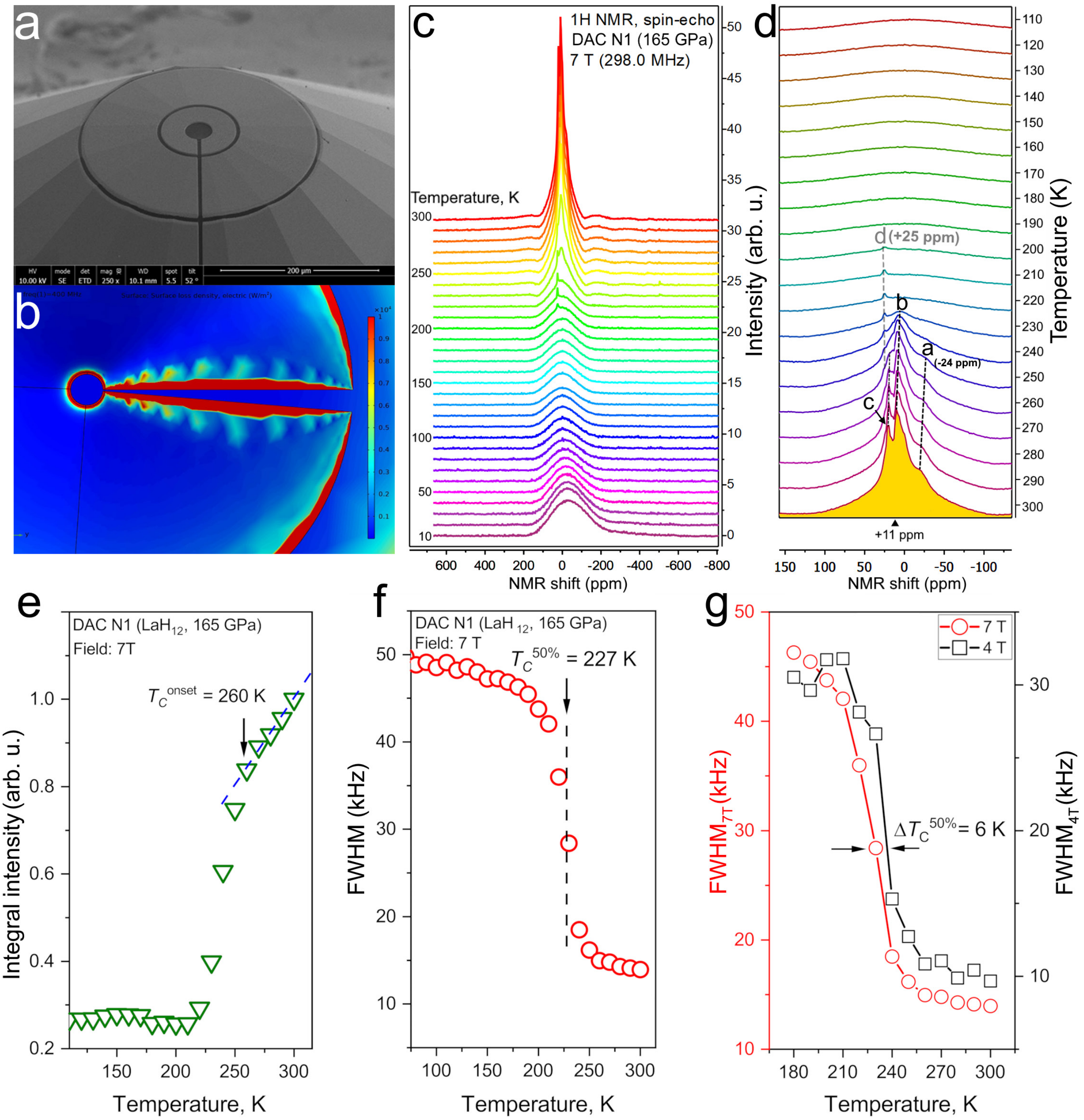

**Figure 3. $^1H$ Nuclear magnetic resonance experiments using DAC N1.** (a) Scanning electron microscopy of the copper Lenz lens of DAC N1, prepared by using a focused Ga-ion beam. (b) Modeling of the distribution of surface current density losses (in $W/m^2$) in the Lenz lens at 400 MHz (using COMSOL Multiphysics[52]). (c)

Stack of $^1$H NMR spectra, recorded during cooling from 300 to 10 K in steps of 10 K. (d) $^1$H NMR spectra in the temperature regime of the superconducting transition. The structure of the signal above 240 K is composed of at least four different peaks ("a-d"). (e) Temperature dependence of the $^1$H NMR signal area ("intensity") between 115 and 300 K. $T_c^{onset}$ = 260 K denotes the onset of the pronounced intensity decrease. (f) Temperature dependence of the $^1$H NMR linewidth (FWHM = full width at half maximum) between 75 and 300 K, yielding $T_c$(50 %) = (227 ± 5) K. (g) Temperature dependence of FWHM for applied magnetic fields of 4 and 7 T. At 4 T, $T_c$ is approximately 6 K higher.

To confirm the onset of SC as origin of the pronounced change in the spectral features of the sample in DAC N1 and to probe the electronic low-energy excitations, we also measured the nuclear spin-lattice relaxation time $T_1$. The relaxation rate $1/T_1T$ of the samples in DACs N1 and N3 begins to decrease at temperatures above 260 K at 7 T (Fig. 4a). A sharp drop in $1/T_1T$ by 3-6 times in the temperature range between 220 and 260 K, not observed in the calibration samples and other DACs, indicates the superconducting nature of the transition in the lanthanum-hydride samples in DACs N1 and N3. This confirms our findings based on the RF transmission studies discussed above (Fig. 1).

For the DACs N1 and N3, the overall behavior of the $^1$H NMR spectra and the spin-lattice relaxation time is very similar. This can be understood by the presence of the same $P6$-$LaH_{12}$ phase in both DACs (Fig. 2).

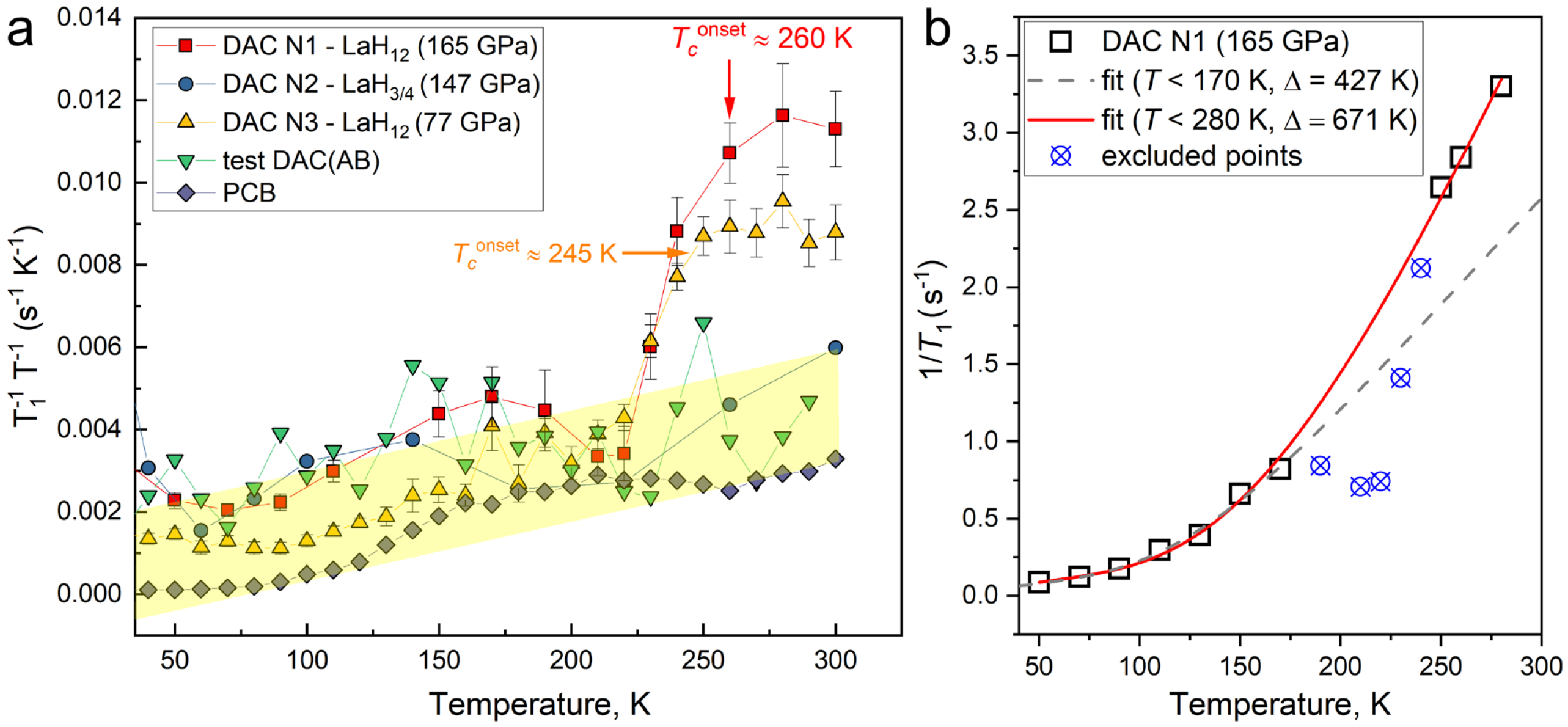

**Figure 4. $^1$H nuclear spin-lattice relaxation rate $(T_1T)^{-1}$ of $LaH_{12}$ (samples in DAC N1 and DAC N3), and in other DACs and test samples.** (a) Temperature dependence of the relaxation rate $1/T_1T$. There is a pronounced drop in the relaxation rate of the samples in DACs N1 and N3 below 260 K and 245 K, respectively. This is associated with the presence of the $LaH_{12}$ phase, which is absent in the other DACs and test samples. (b) Fits using $1/T_1 = Ae^{-\Delta/T} + CT$ (here we set $k_B = 1$) to describe the temperature-dependent $1/T_1$ of the sample in DAC N1 at 165 GPa and 7 T. Here, $A$ and $C$ are fit parameters, as well as the superconducting gap $\Delta = \Delta(0)$. We performed fits for temperatures below 170 and 280 K, indicated by a gray dashed line and a red solid line, respectively.

## Discussion

The complete volume displacement of an applied external magnetic field in the superconducting state is a key characteristic of type-I superconductors. By contrast, in type-II superconductors, in particular for granulated ones[11], the field penetrates the sample volume[53,54,55]. As the NMR signal intensity is proportional to the penetrated sample volume, this facilitates NMR measurements even at temperatures below $T_c$ (Fig. 4a, b). The measured spin-lattice relaxation time of the sample in DAC N1 at 300 K is about 300 ms, and the product is $TT_1 \approx 88$ sK. Comparing the low-temperature behavior of the spin-lattice relaxation of the samples in DACs N1 and N2, we note that the differences in $TT_1$ are small: $TT_1$ = 573 sK at 50 K for the sample in DAC N1, and 340 sK at 40 K for the sample in DAC N2. This is not surprising, given the fact that superconducting phases yield only a very small contribution to the NMR signal at $T << T_c$. The $^1$H signal at low temperatures is mainly due to the residual non-superconducting hydrogen-containing compounds.

As Fig. 4a,b and the Supplementary Fig. S19 show, there is a region of increased spin-lattice relaxation rate between about 100 and 227 K for the sample in DAC N1, which can be associated with either $LaH_{12}$ or residual $NH_3BH_3$. Phenomenologically, this feature in $1/T_1$ vs $T$ resembles the relaxation behavior of $^{27}$Al NMR measurements, studied in the classical Hebel-Slichter[24,25] and Redfield-Anderson[23] experiments (Supplementary Fig. S20). In this scenario, the hump below $T_c^{50\%}$ for the sample in DAC N1 may be caused by the contribution of a Hebel-Slichter-type coherence peak, indicating the opening of the SC gap and the appearance of narrow maxima in the density of quasiparticle states $N(E)$. Since the dependence $1/T_1 \propto N(E)^2$ is quadratic, the presence of a maximum in $N(E)$ leads to a peak in $1/T_1$ (the Hebel-Slichter peak). If this increase of the relaxation rate is indeed related to such a coherence peak and not to a residual amount of $NH_3BH_3$ (Supporting Fig. S9), it would be a strong argument in favor of conventional electron-phonon pairing in lanthanum superhydrides[56,57].

However, we would like to emphasize another observation, based on our $T_1$ data. In many superconductors, antiferromagnetic fluctuations[58], as well as a strong electron-phonon interaction[59], may suppress a possible Hebel-Slichter peak. $LaH_{12}$ is expected to be a superconductor in the strong-coupling limit. Indeed, describing the temperature dependence of $1/T_1$ below $T_c$ using an exponential fit allows to determine the superconducting gap of $LaH_{12}$ at 165 GPa. As shown in Fig. 4b (and Fig. S19), such fits to the $1/T_1$ data for different temperature regimes provide an estimate of the superconducting gap $\Delta(0)$. We obtain values between about 427 and 671 K (36.8 and 57.8 meV), yielding $R_\Delta = 2\Delta(0)/k_BT_c$ of about 3.76 to 5.16, in qualitative agreement with the strong electron-phonon coupling extension of the BCS theory proposed by Migdal and Eliashberg[60,61].

The onset temperature of the superconducting transition in $LaH_{12}$ deserves a separate discussion. Although even in pure $Fm\overline{3}m$-$LaH_{10}$ in zero field, the critical temperature $T_c^{onset}$ does not exceed about 250 K, in our experiment, we observe an unusual behavior of the NMR spectra and $1/T_1T$ right above 260 K at 7 T. Considering a typical value of the derivative $dB_{c2}/dT \approx -1$ T/K for superhydrides[62,63], we estimate that at zero field, SC in the sample in DAC N1 sets in already around 267 K, which is significantly higher than the $T_c^{onset}$ observed in electrical-transport measurements[2]. Thus, the results of our NMR study in magnetic fields of 4 and 7 T agree well with the RF transmission measurements in zero field (Fig. 1).

Measurements of the electrical transport are limited to the current path of least resistance. By contrast, RF methods sense the full bulk of the sample. Hence, differences in $T_c^{onset}$ detected by these techniques are a natural consequence. Once SC sets in, the resistance drops to zero and any information below that transition is hidden. In inhomogeneous granular superconductors consisting of multiple phases with varying $T_c$, a series-connected normal-metallic and superconducting resistance may be sensed. Hence, step-like transitions with non-zero resistances may be detected. Indeed, such features have been noted previously in La-H[3,41,43] and La-Sc-H systems[12] at temperatures much higher than 250 K. It is interesting to note that similar anomalies were also observed in the high-frequency AC-susceptibility measurements on $LaH_{10\pm x}$ samples[42].

## Conclusion

Employing $^1$H nuclear magnetic resonance spectroscopy and radio-frequency transmission measurements, we studied five lanthanum-hydride samples, including two crystal modifications of $LaH_{12}$, $LaH_{9+x}$, tetragonal $LaH_4$, and $LaH_3$, at various pressures from 19 to 165 GPa in magnetic fields of 0, 4, and 7 T. The superconducting transition in the newly discovered $P6$-$LaH_{12}$ is characterized by a strong suppression of the $^1$H-NMR signal intensity below about 260 K in a magnetic field of 7 T. Furthermore, it is accompanied by a pronounced change in all NMR characteristics, including the spin-lattice relaxation rate $1/T_1T$, which exponentially decreases at temperatures below $T_c$. Our estimate of the superconducting gap in $LaH_{12}$ gives $\Delta(0)$ of about 427 to 671 K (corresponding to 36.8 to 57.8 meV) and $R_\Delta = 2\Delta(0)/k_BT_c$ of about 3.76 to 5.16, in accordance with the strong-coupling scenario in lanthanum superhydrides.

We confirmed these observations by additional radio-frequency transmission experiments on the same samples using a transformer configuration, where Lenz lenses act as primary and secondary coils. The transmission amplitude exhibits abrupt changes, already setting in at (267 ± 5) K (in zero field). Remarkably, the La-H system contains phases with significantly higher $T_c$ than what was previously reported based on electrical-transport measurements. In the two samples in DACs N1 and N3, the detected changes in the $^1$H-NMR spectral characteristics, and, most importantly, the spin-lattice relaxation rate $1/T_1T$, start about 15-20 K above the zero-field onset $T_c$ of the canonical $LaH_{10}$, determined by means of electrical-transport measurements.

# Methods

## 1. *Diamond-anvil cells*

To prepare the DACs N1 and N2, 300 μm thick tungsten gaskets were used. These were compressed to a pressure of 20 GPa before drilling holes that were 50-100 % of the diameter of the diamond anvil culet. After drilling the holes, the surface of the tungsten gaskets was coated on both sides with a 1-2 μm thick layer of tantalum, using magnetron. We then heated the gaskets in air to 1000 °C to oxidize Ta and form $Ta_2O_5$. The resulting W/Ta/$Ta_2O_5$ insulating gasket with a layer of $Ta_2O_5$ (contact resistance is above 6 MΩ) is reliably insulated tungsten gasket from Lenz lenses deposited on the diamond anvils. DACs N3 and N4 were prepared using rhenium (Re, 300 μm thick) gaskets with a sputtered insulating layer of ~50 nm thick aluminum oxide. The aluminum oxide was prepared by DC magnetron sputtering of aluminum in an oxygen atmosphere.

For the DACs N1-N4, Lenz lenses made of deposited Cu/Mo and Cu layers with a thickness of 0.6-2 μm were fabricated. Magnetron sputtering was carried out in DC mode. The total thickness of the metal layer was 600-755 nm (Fig. S1). The width of the ion etching zones separating the conducting rings of the Lenz lenses was about 5 μm. Three-stage Lenz lenses, with a diameter of the annular etching zones of 60 μm, 300 μm and 3 mm (Fig. S1c) were used.

The electrical circuit of all DACs included an RF emitter in the form of a single-turn coil on a 0.3 mm thick printed circuit board (PCB) made out of textolite or Teflon, 0.1 mm thick Cu or Au conductive patterns with a diameter of 3-4 mm, depending on the diameter of the diamond anvil. The RF coil was equipped with a 5 nF ceramic matching capacitor. The RF emitters were connected in series to form the contour of a two-turn coil. Organic glue was used in minimal quantities and as far away from electrical contacts as possible.

## 2. *$LaH_x$ samples*

The samples in DAC N0, N1 and N2 were synthesized using laser heating of metallic La microparticles and $NH_3BH_3$, while the samples in N3 and N4 were obtained from $LaH_{3-x}$ and $NH_3BH_3$. The infrared (1.04 μm) laser heating under high pressure consisted of a series of 10-20 pulses with a duration of 0.1-0.3 seconds. Temperature was not measured during laser heating.

## 3. *Nuclear magnetic resonance*

The $^1$H NMR experiments were conducted in a superconducting magnet, which reaches magnetic fields up to 8 T. The radial homogeneity of the field is better than 2.5 ppm over 10 mm, whereas the longitudinal field varies less than 4 ppm within ± 5 mm from the center of the field. The magnet is equipped with a $^4$He flow cryostat, which can reach temperatures from 1.6 to 300 K, with temperature stabilization of much better than 1 %. All NMR data were recorded using a commercial, phase-coherent spectrometer with a maximum frequency of 500 MHz, and an RF power amplifier, achieving a pulse power of up to 250 W. Standard Hahn-spin-echo sequences were used to measure the $^1$H NMR spectra. The pulse sequences were typically spaced apart by delay times of typically 500 ms and we usually accumulated 2000 scans for a $^1$H spectral signal-to-noise ratio between 80 and

300. To process the NMR data, we used a home-built LabView software and the MestReNova program[64].

*4. Radio-frequency transmission*

Measurements of the radio-frequency transmission were performed through the Lenz-lens system in the DACs in a circuit with two lock-in amplifiers (Fig. S2). An SR844 lock-in amplifier (Stanford Research) was used as high-frequency signal generator, as well as a receiver to measure the transmitted signal at the same frequency. The sample was placed in a low-frequency magnetic field (19 Hz or 33 Hz, maximum induction between 20 and 40 Gauss), generated by a solenoid. The current feeding the solenoid was created by an SR830 lock-in amplifier, working in tandem with a Yamaha PX3 power amplifier. The current amplitude in the solenoid circuit was typically 0.5–5 A. The envelope of the high-frequency signal near the superconducting transition temperature contains even harmonics, of which the strongest, the 2$^{nd}$ harmonic, was detected.

*5. X-ray diffraction*

The ambient temperature X-ray diffraction measurements were performed using the synchrotron X-ray diffraction on the ID27 beamline with a beam of wavelength of 0.3738 Å at the European Synchrotron Research Facility (ESRF, proposal MA-5924).

Mapping of the sample center was carried out on a 10×10 and 4×4 grids with a step of 5 microns, accumulating time was about 10 seconds per image. The experimental XRD images were integrated and analyzed using the Dioptas 0.6 software package [65]. To fit the diffraction patterns and obtain the cell parameter, we analyzed the data using Mercury 2021.2.0 [44] and Jana2006 software [66], employing the Le Bail method [67]. Analysis of the spatial distribution of hydride phases in the sample was performed using the XDI [45] and Dioptas 0.6 programs.

*6. SEM/FIB*

SEM images and FIB assisted lithography for the fabrication of Lenz lenses were done with the help of Helios DualBeam systems (Xenon plasma and Gallium ions) from ThermoFisher Scientific.

## Acknowledgments

D. V. S. and D. Z. thank the National Natural Science Foundation of China (NSFC, grant No. 12350410354) and the Beijing Natural Science Foundation (grant No. IS23017) for the support of this research. D. Z. thanks the China Postdoctoral Science Foundation (No. 2023M740204) and the Fundamental Research Funds for the Central Universities for support of this research. D.S. and D.Z. are grateful for financial support from the EMFL-ISABEL project. This work was supported by the National Key Research and Development Program of China (grant 2023YFA1608900, subproject 2023YFA1608903). V.V.S. acknowledges financial support from the Shanghai Science and Technology Committee, China (No. 22JC1410300) and Shanghai Key Laboratory of Materials Frontier Research in Extreme Environments (MFree), China (No. 22dz2260800). I.A.T. acknowledges the support by the state assignment of the Kurchatov Complex of Crystallography and Photonics (KKKiF). We express our gratitude to Dr. Fuyang Liu, Dr. Yanping Yang and Dr. Hongliang Dong from Beijing and Shanghai HPSTAR for their help with the Ga FIB ion etching and magnetron sputtering of $Al_2O_3$, respectively. We also thank Dr. Thomas Meier (HPSTAR, Bejing) for technical support and drawing our attention to the NMR research using diamond anvil cells. We thank Prof. Yang Ding's group, especially Dr. Jianbo Zhang, at HPSTAR, Beijing, for allowing us to use the Raman spectrometer and the magnetron sputtering system. We would like to express our gratitude to Dr. Liuxiang Yang (HPSTAR, Beijing) for the laser heating of our samples. We thank

Dr. Xiaoli Huang (Jilin University) and Dr. Liling Sun (HPSTAR, Beijing) for assistance with X-ray diffraction studies at the Shanghai synchrotron research facility. We also express our gratitude to the teams of the high-pressure stations ID27 (ESRF) and BL15U1 (SSRF) for their help in the XRD investigations. We acknowledge support from the Deutsche Forschungsgemeinschaft (DFG) through SFB 1143 (Project No. 247310070) and the Würzburg-Dresden Cluster of Excellence on Complexity and Topology in Quantum Matter - ct.qmat (EXC 2147, Project No. 390858490), as well as the support of the HLD at HZDR, member of the European Magnetic Field Laboratory (EMFL). We thank the Department for Physics of Quantum Materials at the MPI CPfS (Dresden, Germany) and Ion Beam Center at HZDR (Dresden, Germany) for putting their FIB infrastructure at our disposal.

## Author contributions

D.V.S., D.Z. and I.A.T. designed the experiments, prepared the DACs including sputtering of Lenz lenses and synthesis of the lanthanum hydride samples, did the COMSOL modelling, performed the RF transmission measurements, transport, powder X-ray diffraction studies, and analyzed the XRD data. F.B., D.V.S., D.Z., and S.L. performed the NMR measurements and analyzed the NMR data. J.W., H.K., and T.H. supervised the work at HZDR. V.V.S. supervised the work at HPSTAR. All authors contributed to the discussion of the results and the preparation of the manuscript.

*Dedicated to the memory of Dr. Mikhail I. Eremets*

SUPPORTING INFORMATION

# Transmission of radio-frequency waves and nuclear magnetic resonance in lanthanum superhydrides

Dmitrii V. Semenok [1,†,*], Florian Bärtl [2,3,†], Di Zhou [1,†], Toni Helm[2,4], Sven Luther[2], J. Wosnitza[2,3], Ivan A. Troyan[5], Viktor V. Struzhkin[1,6,*], and Hannes Kühne[2,*]

[1] *Center for High Pressure Science and Technology Advanced Research (HPSTAR), Beijing*

[2] *Dresden High Magnetic Field Laboratory (HLD-EMFL) and Würzburg-Dresden Cluster of Excellence ct.qmat, Helmholtz-Zentrum Dresden-Rossendorf, 01328 Dresden, Germany*

[3] *Institute for Solid State and Materials Physics, Technische Universität Dresden, 01062 Dresden, Germany*

[4] *Max Planck Institute for Chemical Physics of Solids, 01187 Dresden, Germany*

[5] *A.V. Shubnikov Institute of Crystallography of the Kurchatov Complex of Crystallography and Photonics (KKKiF), 59 Leninsky Prospekt, Moscow 119333, Russia*

[6] *Shanghai Key Laboratory of Material Frontiers Research in Extreme Environments (MFree), Shanghai Advanced Research in Physical Sciences (SHARPS), Pudong, Shanghai 201203, China*

*Corresponding authors, emails: dmitrii.semenok@hpstar.ac.cn (Dmitrii V. Semenok), viktor.struzhkin@hpstar.ac.cn (Viktor V. Struzhkin), h.kuehne@hzdr.de (Hannes Kühne).

†These authors contributed equally to this work

## Content

## 1. Experimental

For preparation of the DACs N1 and N2, we used 300 μm thick tungsten (W) gaskets, compressed to a pressure of 20 GPa before drilling holes that were 50-100 % of the diameter of the diamond anvil culet. After drilling the holes, we coated the surface of the tungsten gaskets on both sides with a 1-2 μm thick layer of tantalum using magnetron sputtering for 30 minutes at a DC power of 150-200 W. We then heated the gaskets in air to 1000 °C for 5 seconds using an open flame to oxidize Ta and form $Ta_2O_5$. Due to the very good adhesion between the layers in the W/Ta/$Ta_2O_5$ interface, the resulting insulating gasket with a layer of $Ta_2O_5$ (contact resistance is above 6 MΩ) is reliably insulated from Lenz lenses deposited on the diamond anvils.

DACs N3 and N4 were prepared using rhenium (Re, 300 μm thick) gaskets with a sputtered insulating layer of ~50 nm thick aluminum oxide. Aluminum oxide was prepared by DC magnetron sputtering of aluminum in an oxygen atmosphere. The adhesion of such a layer to the Re gasket is very weak and any mechanical impact may destroy the insulating layer. This makes it inconvenient for the practical use.

For the DACs N1-N4, we fabricated Lenz lenses made of deposited Cu/Mo and Cu layers with a thickness of 0.6-2 μm. We tested the use of a two-component Mo/Cu layer, where molybdenum was sputtered to increase the adhesion of the metal layer to the diamond surface. Magnetron sputtering was carried out in direct-current mode, the molybdenum layer was sputtered for 5 minutes with a power of 150 W, while the subsequent copper layer was sputtered for 45 minutes with a power of 75 W. The total thickness of the metal layer was 600-755 nm (Figure S1). The width of the ion etching zones separating the conducting rings of the Lenz lenses was about 5 μm. We used three-stage Lenz lenses, with a diameter of the annular etching zones of 60 μm, 300 μm and 3 mm (Figure S1c). Lenz lenses are fairly defect-resistant structures. Despite point-like damage during pressure build-up, they retain their functionality. This is unlike four-electrode wirings for transport measurements, where even a small damage of any electrode, a few microns wide, can be fatal for the experiment.

The electrical circuit of all DACs included an RF emitter in the form of a single-turn coil on a 0.3 mm thick printed circuit board (PCB) made out of textolite or Teflon, 0.1 mm thick Cu or Au conductive patterns with a diameter of 3-4 mm depending on the diameter of the diamond anvil. We also tested ceramic ($Al_2O_3$) printed circuit boards, but they were too brittle and almost always broke when the DACs were pressurized. The RF coil was equipped with a 5 nF ceramic matching capacitor. We realized that in many cases it is more convenient to work without using this capacitor, especially when measuring RF transmission. The RF emitters were glued to WC, BN, or BeCu seats using inorganic H-free glue made from a solution of sodium silicate in deuterated water, and connected in series to form the contour of a two-turn coil. Organic glue was used in minimal quantities and as far away from electrical contacts as possible.

**Table S1.** Parameters of samples, high-pressure DACs, Lenz lenses, gaskets and diamond anvils used in this study. “sc” means simple cubic space group ($Pm\bar{3}m$), “h” means “hexagonal”. WC – is the nonmagnetic tungsten carbide, and c-BN is the cubic boron nitride.

| | **DAC N0** | **DAC N1** | **DAC N2** | **DAC N3** | **DAC N4** |
|---|---|---|---|---|---|
| **Starting material** | La/AB | La/AB | La/AB | $LaH_{3-x}$/AB | $LaH_{3-x}$/AB |
| **DAC’s material** | BeCu | 40HNU | BeCu | BeCu | BeCu |
| **PCB material** | teflon | teflon | teflon | textolite | textolite |
| **Gasket** | nonmagnetic steel/MgO | W/Ta/$Ta_2O_5$ | W/Ta/$Ta_2O_5$ | Re/$Al_2O_3$ | Re/$Al_2O_3$ |
| **Culet diameter, μm** | 80 | 75 | 100 | 100 | 100 |
| **Glue** | Epoxy | $Na_2SiO_3 \times D_2O$ | $Na_2SiO_3 \times D_2O$ | Superglue | Superglue |
| **Lenz lens** | Ta/Au, physical mask | Cu / Ga FIB | Cu/ lithography | Cu/Mo, Ga FIB | Cu, Ga FIB |
| **Seat material** | Non-magnetic WC | Non-magnetic WC | Non-magnetic c-BN | Non-magnetic WC | Non-magnetic WC |
| **Pressure, GPa** | 149-120 | 165 | 147-150 | 80-90 | 19 |
| **Composition according to XRD** | N/A | h-$LaH_{12}$ | $LaH_3$ + $LaH_4$ | sc-$LaH_{12}$ + h-$LaH_{12}$ | $LaH_2$+$LaH_3$ |
| **Signal/Noise ratio in non-SC state (number of scans)** | N/A | 270 ($10^4$) | 144 ($2\cdot10^3$) | 560 ($2\cdot10^3$) | 15 ($1.3\cdot10^5$) |

We conducted all NMR experiments in a superconducting magnet, which reaches magnetic fields up to 8 T. The radial homogeneity of the field is better than 2.5 ppm over 10 mm, whereas the longitudinal field varies less than 4 ppm within ±5 mm from the center of the field. The magnet is equipped with a $^4$He flow cryostat, which can reach temperatures from 1.6 to 300 K, with temperature stabilization of much better than 1 %. We recorded all NMR data using a commercial, phase-coherent spectrometer with a maximum frequency of 500 MHz, and an RF amplifier, achieving pulse power of up to 250 W. We used standard Hahn-spin-echo sequences (π/2 → τ → π) to measure the $^1$H NMR spectra. The pulse sequences were typically spaced apart by delay times in the range of 500 ms and we usually accumulated 2000 scans for a spectral signal-to-noise ratio between 80 and 300. To process the NMR spectra and construct graphs, we used a home-built LabView software and the MestReNova 12.0.0 program[1].

In order to obtain a very stable resonance frequency of the NMR-resonant circuit, we chose a top-tuned configuration, where we placed the capacitors of the resonant circuit outside of the cryostat and operated them with stepper motors for automated tuning and matching. Additionally, we used a phase shifter to achieve a wide tuning range of the resonance frequency. This setup enables a broadband frequency tuning of the resonant circuit and guarantees stable resonance conditions for the whole temperature range.

The spin-lattice-relaxation rate was measured using a saturation-recovery sequence (π/2 → τ’ → π/2 → τ → π). For short values of τ’, the NMR signal intensity (= $M_z$) is close to zero, whereas for long values of τ’, the intensity saturates. We evaluated the resulting relaxation curve by fitting a single-exponential function for spin-1/2 nuclei to our data according to

$$M_z(\tau) = M_{z0}\left[1 - e^{-\left(\frac{\tau'}{T_1}\right)^{\beta}}\right] + offset, \qquad \text{(S1)}$$

where $M_{z0}$ is the equilibrium nuclear magnetization, $offset$ is a parameter that accounts for non-perfect nutation conditions during the NMR pulse sequence (and is very small compared to the contribution of the exponential term in our experiments). The stretching exponent $0 < \beta \leq 1$ is a parameter that reflects the distribution width of $T_1$. If $\beta = 1$, every excited nuclear spin relaxes with the same $T_1$. For smaller $\beta$, the underlying distribution function of $T_1$ broadens.

## 2. *Preparation of Lenz lenses*

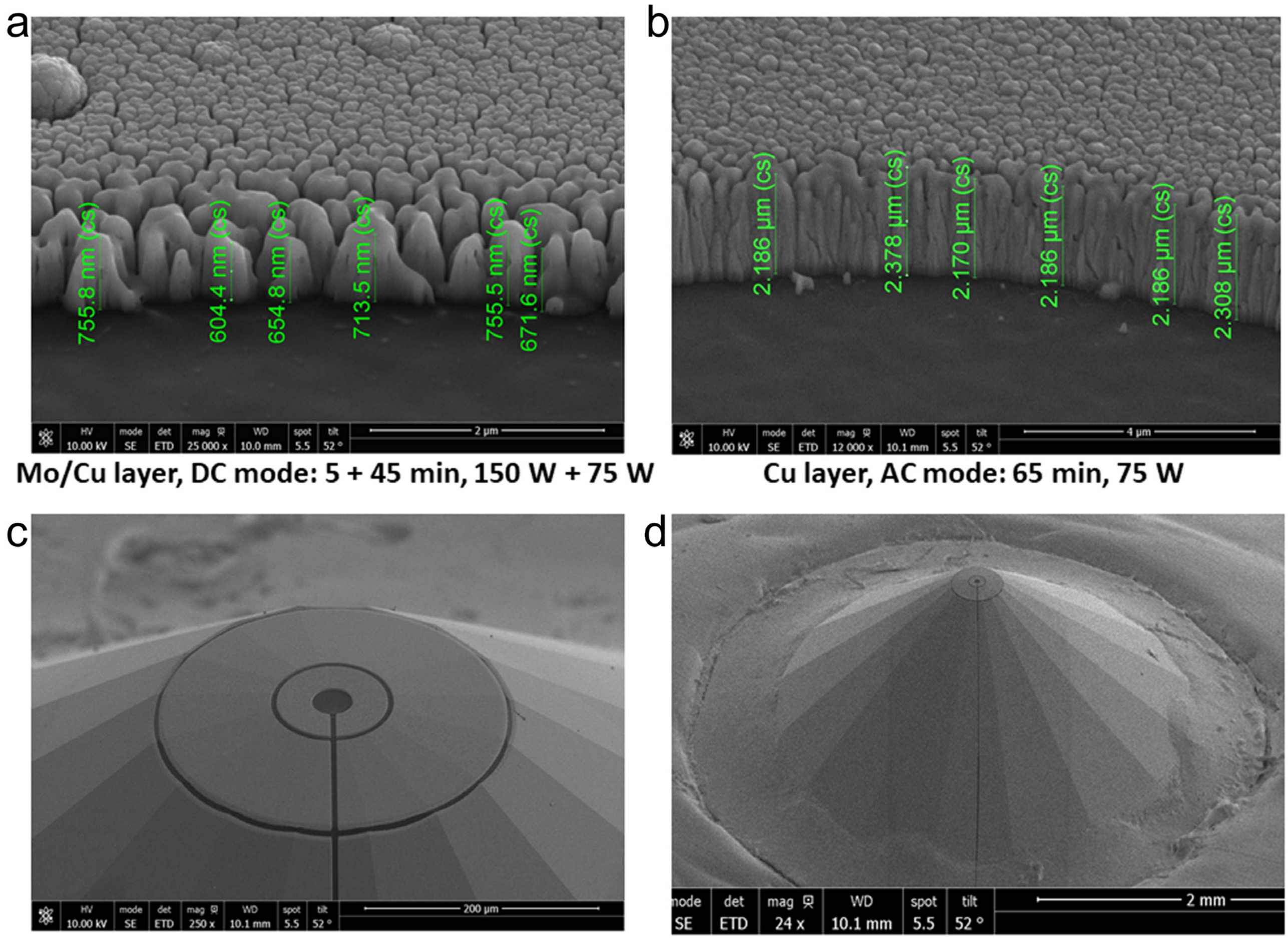

**Figure S1.** Scanning electron microscopy of a typical result of magnetron sputtering followed by ion etching of Lenz lenses on a diamond anvil. (a) Determination of the thickness of the two-layer Mo/Cu coating. (b) Determination of the thickness of the sputtered pure copper film. Sputtering pure copper was carried out in alternating-current mode for 65 minutes at a power source of 75 W. The layer thickness was more than 2 microns. (c) Electron microscopy of a diamond anvil with a sputtered Lenz lens. (d) Continuation of the radial cut to the boundaries of the sprayed metal layer.

*3. Radio-frequency transmission measurements*

We performed measurements of the transmission of high-frequency signals through the Lenz-lens system in the DACs in a circuit with two lock-in amplifiers (Figure S2). We used a SR844 lock-in amplifier (Stanford Research) as high-frequency signal generator, as well as receiver to measure the transmitted signal at the same frequency. We placed the sample in a low-frequency magnetic field (19 Hz or 33 Hz, maximum induction between 20 and 40 Gauss), generated by a special solenoid. The current feeding the solenoid was created by an SR830 lock-in amplifier, working in tandem with a Yamaha PX3 power amplifier. The current amplitude in the solenoid circuit is typically 0.5–5 A. The envelope of the high-frequency signal near the superconducting transition temperature contains even harmonics, of which the strongest, the 2$^{nd}$ harmonic, was detected, thus serving as an additional marker of the superconducting transition. In high-frequency measurements, the target signal can have two basic forms: (1) A peak, hump, or trough of various shapes; (2) a step. Typical measurement parameters are given in Table S2.

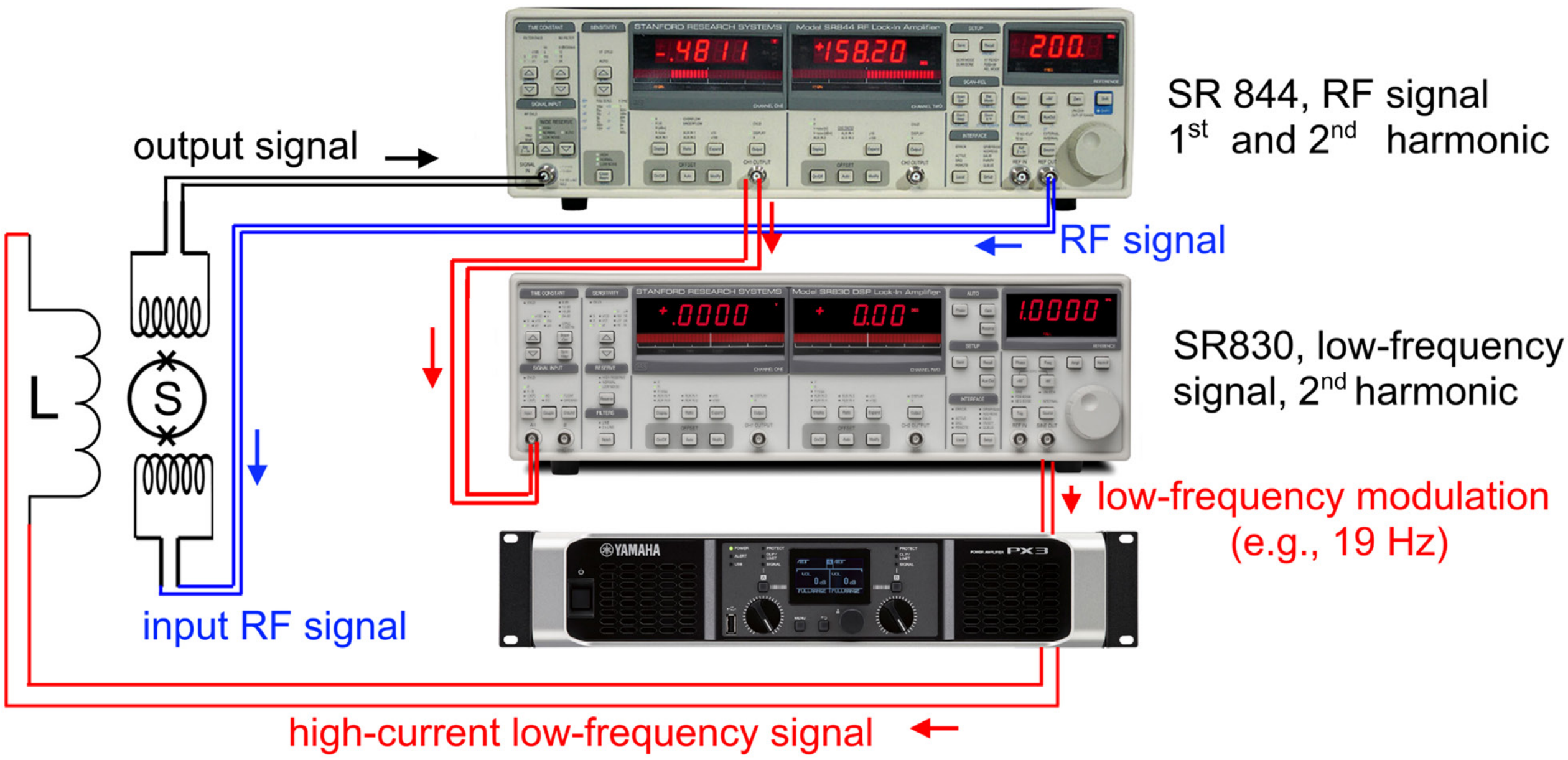

**Figure S2.** Scheme of the setup for radio-frequency transmission studies of the samples in the DACs. "L" represents an external solenoid creating low-frequency modulating magnetic field (usually 19 or 33 Hz). "S" stands for sample.

**Table S2.** Parameter settings for the SR844 and SR830 lock-in amplifiers used in our RF measurements.

| | **SR844** | **SR830** |
|---|---|---|
| **Time constant** | 100 µs | 300 ms |
| **Filter** | 18 or 24 dB | 18 or 24 dB |
| **Signal input** | 50 Ω (without capacitors)<br>30 pF (with capacitors) | AC |
| **Wide Reserve** | Low noise or Normal | Low noise or Normal |
| **Harm #** | 1 or 2 | 2, 4, 6 … |

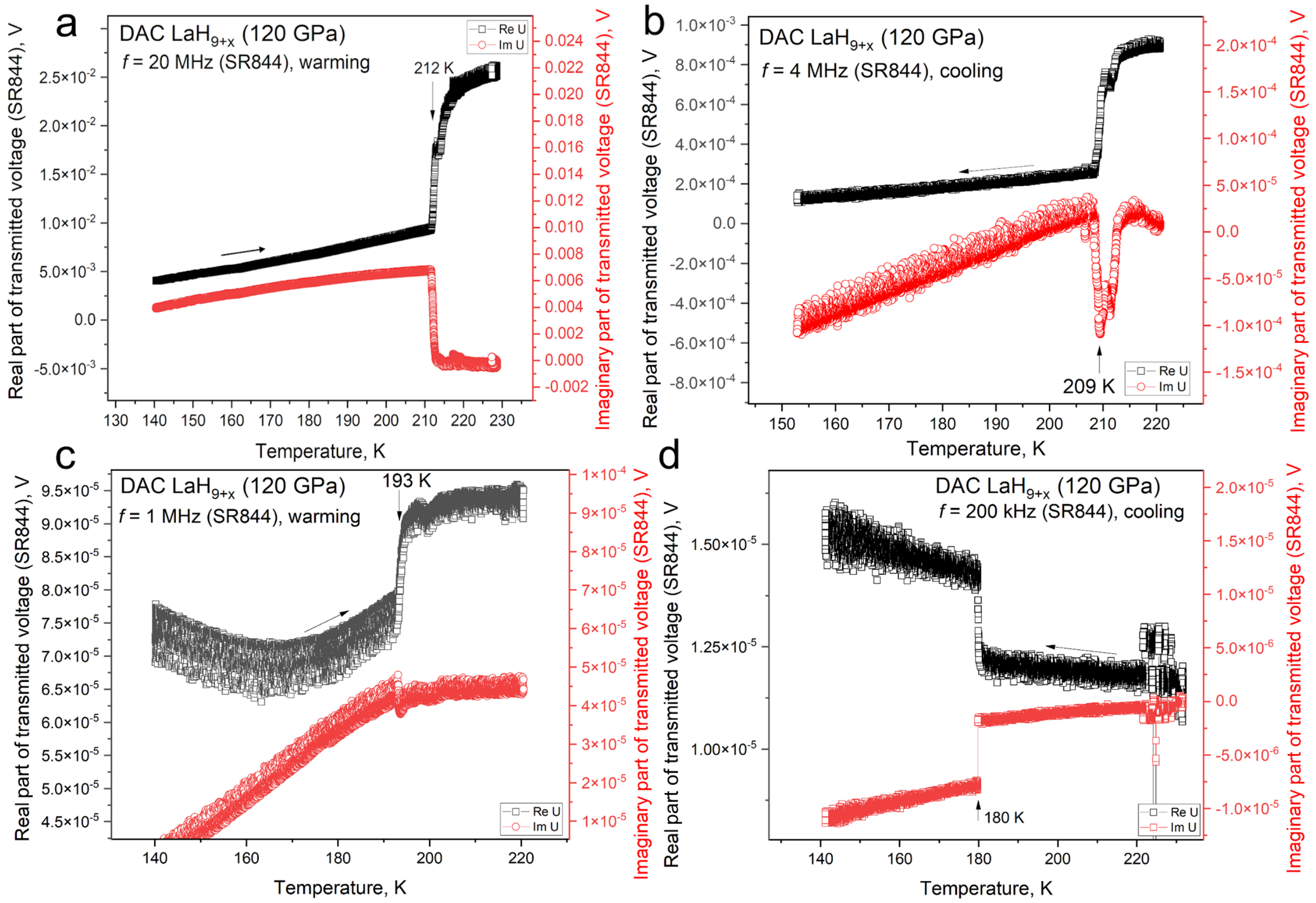

**Figure S3.** Radio-frequency transmission through the sample in DAC N0 ($LaH_{9+x}$) at 120 GPa at different carrier frequencies *f* in cooling and warming cycles. One of the van der Pauw electrode pairs (see Fig. 1 in the main test) was used as an emitting antenna. Real and imaginary components of the transmitted signal at: (a) *f* = 20 MHz. The superconducting transition begins at about 212 K; (b) *f* = 4 MHz; (c) *f* = 1 MHz; (d) *f* = 200 kHz. As the frequency decreases, the hydride phase that is most abundant in the sample has the greatest influence on the RF transmittance.

Features of the passage of radio-frequency electromagnetic waves through a sample can serve as a marker of superconductivity. At the moment of superconducting transition, undamped eddy currents arise on the sample surface, which change the amplitude and phase of the passing RF signal, screening and dissipating it. In addition, the movement of the Abrikosov vortices at the stage of vortex liquid leads to noticeable dissipation peak of the RF waves.

As can be seen in Figures S4 and S5, the features of the radio-frequency transmission and of the second harmonic of the low-frequency magnetic field passing through the sample in DAC N1 are concentrated in two temperature intervals: (1) 248-252 K (Figure S5), which corresponds well to the superconducting properties of cubic $LaH_{10}$ at 165 GPa[2]; (2) the region of 267-286 K (Figure S4), where the signal anomalies are more pronounced. This temperature region corresponds to the onset of the superconducting transition in a previously unknown lanthanide-hydride phase, which dominates the sample, probably *P*6-$LaH_{12}$. It is important to note that resistive transitions at such high temperatures (265-280 K) have been observed in the La-H system by various authors since 2019 [3-5].

The superconducting state in the vicinity of the critical temperature is sensitive to the application of even a weak oscillating (ν) magnetic field. Periodic suppression of the superconducting state leads to modulation of the radio-frequency transmission and the appearance of a second low-frequency harmonic (2ν), which can be observed in our experiments in Figure 4b,c.

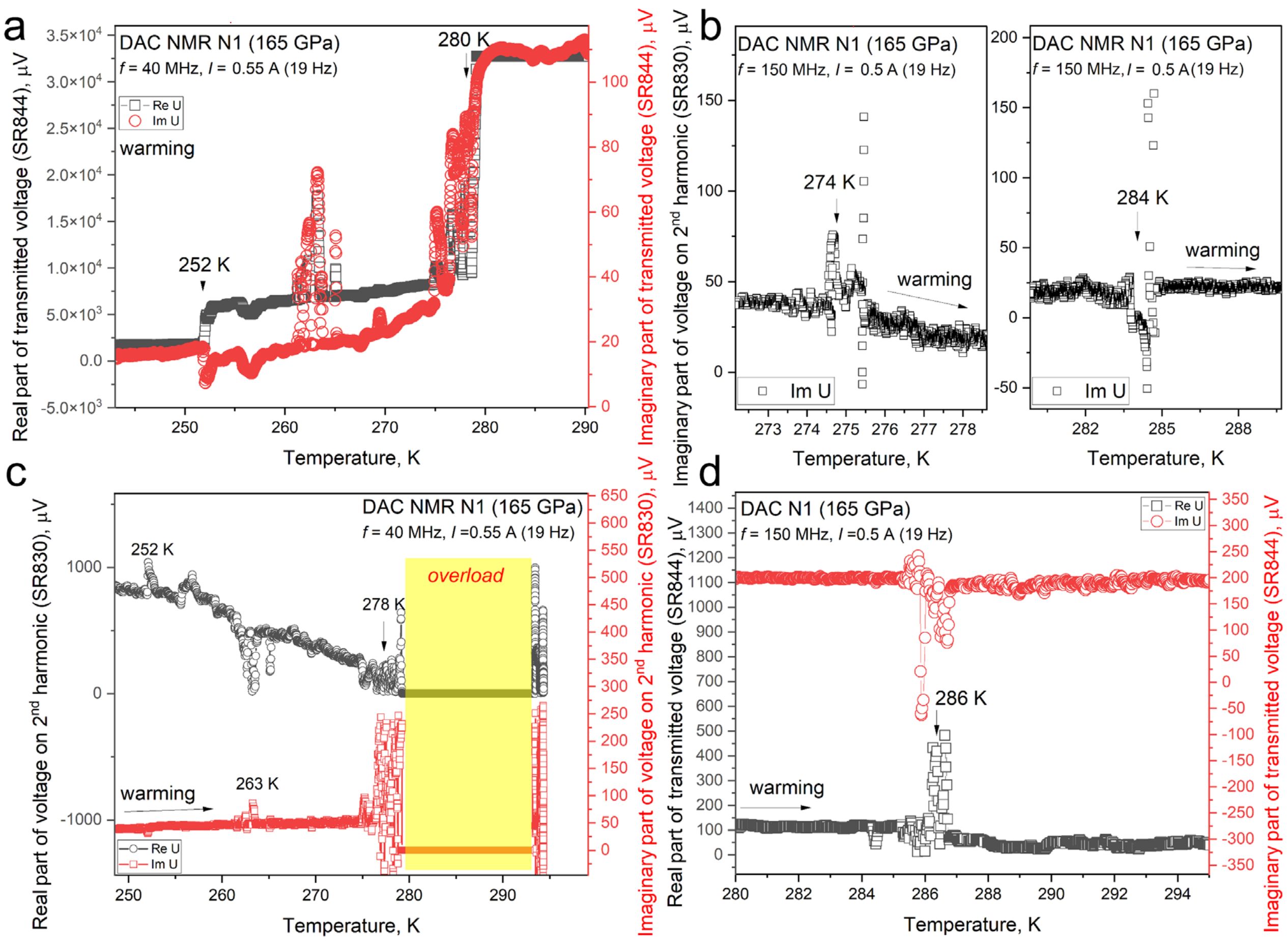

**Figure S4.** Radio-frequency transmission through the sample in DAC N1 (NMR) at 165 GPa at different carrier frequencies $f$ and currents $I$ = 0.5-0.55 A through the external solenoid. (a) Real and imaginary components of the transmitted signal at 40 MHz with a main transition at $T_c \approx$ 275-280 K. The small step at 252 K is likely related to superconductivity of $Fm\bar{3}m$-$LaH_{10}$. (b) Imaginary part of the second-harmonic signal of the low-frequency modulating field (19 Hz) with features around 274 (run 1) and 284 K (run 2). Carrier frequency is 150 MHz. (c) Real and imaginary components of the second-harmonic signal of the low-frequency modulating field (19 Hz). Carrier frequency is 40 MHz. (d) Real and imaginary components of the transmitted signal at 150 MHz (run 2).

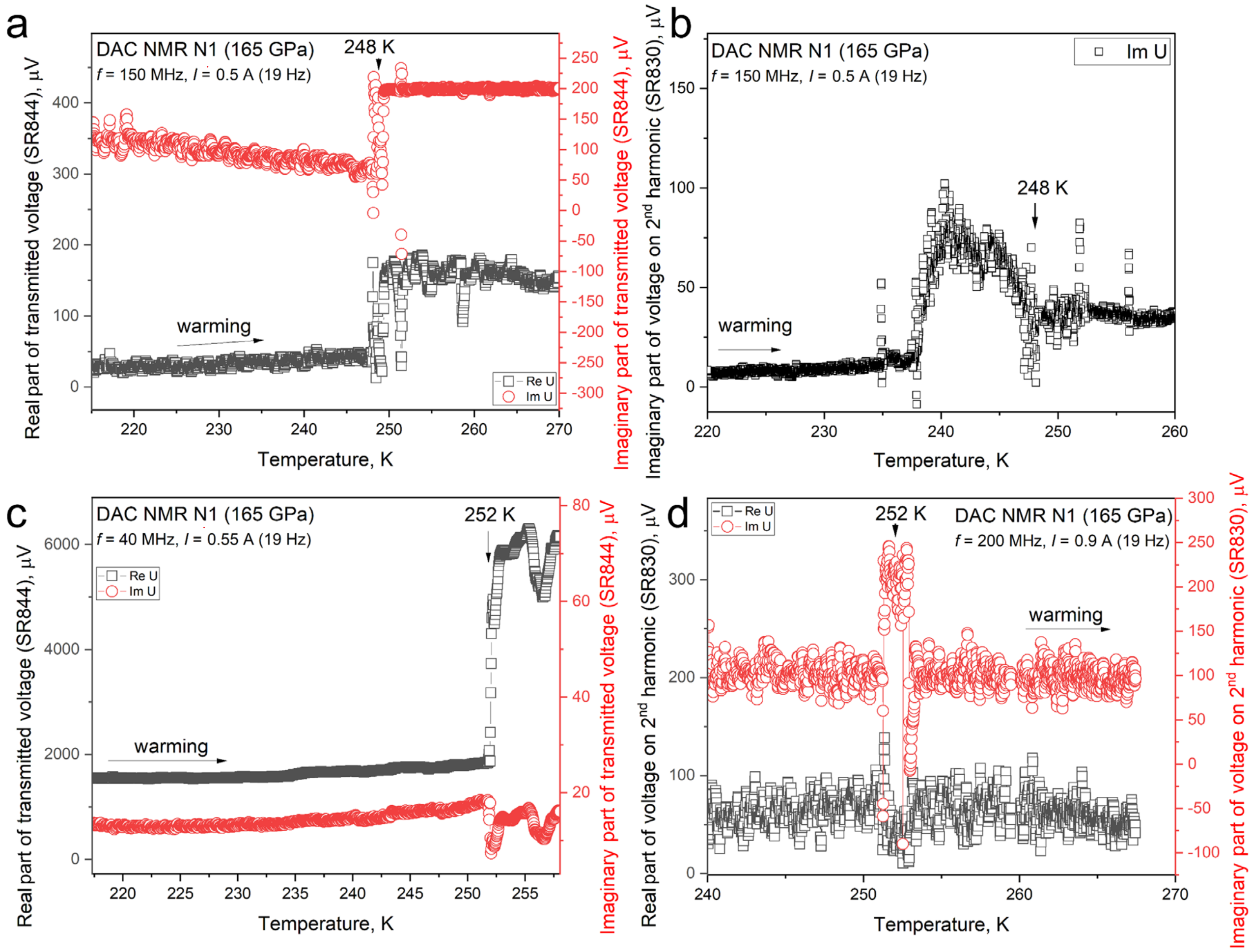

**Figure S5.** Radio-frequency transmission through the sample in DAC N1 (NMR) at 165 GPa at different carrier frequencies *f* around 250 K, far from the region of the high-temperature superconducting transition (260-280 K). (a) Real and imaginary components of the transmitted signal at 150 MHz, with a transition probably corresponding to $Fm\overline{3}m$-$LaH_{10}$. (b) Signal of the second harmonic of the low-frequency modulating field (19 Hz). The onset of the signal at 248 K corresponds to superconductivity in $Fm\overline{3}m$-$LaH_{10}$. (c) Real and imaginary components of the transmitted signal at 40 MHz. (d) Signal of the second harmonic of the low-frequency modulating field (19 Hz) at a carrier frequency of 200 MHz. The pronounced peak probably corresponds to superconductivity in $Fm\overline{3}m$-$LaH_{10}$.

## 4. NMR experiments

In order to validate our NMR measurements, shown in Fig. 3 and 4 of the main text, in particular to demonstrate that we can indeed detect an NMR signal from inside the DAC, we conducted test measurements of a DAC loaded with ammonia borane (AB or $NH_3BH_3$) and an aluminum ($^{27}Al$) foil. In Figure S6c, we show the integrated spectral intensity as function of the interpulse delay time (τ) of a saturation-recovery $T_1$ measurement at 10 K and compare it to a measurement of pure $^{27}Al$ powder, measured in a separate, conventional NMR experiment. We determine $T_1$ and the stretching exponent $\beta$ the same way as described in section 1 (S1). The $T_1$ value of the conventional NMR measurement in the 3He system (182 ms) agrees well with the established Korringa value of $^{27}Al$ (1.85 sK). Our sample in the DAC yields a $T_1$ value that is three times larger (574 ms) by comparison. The reason for this is a non-uniform spatial distribution of the radio-frequency field across the Al foil (stemming from a non-ideal Lenz lens geometry in this experiment), which is also indicated by the low stretching exponent of 0.4.

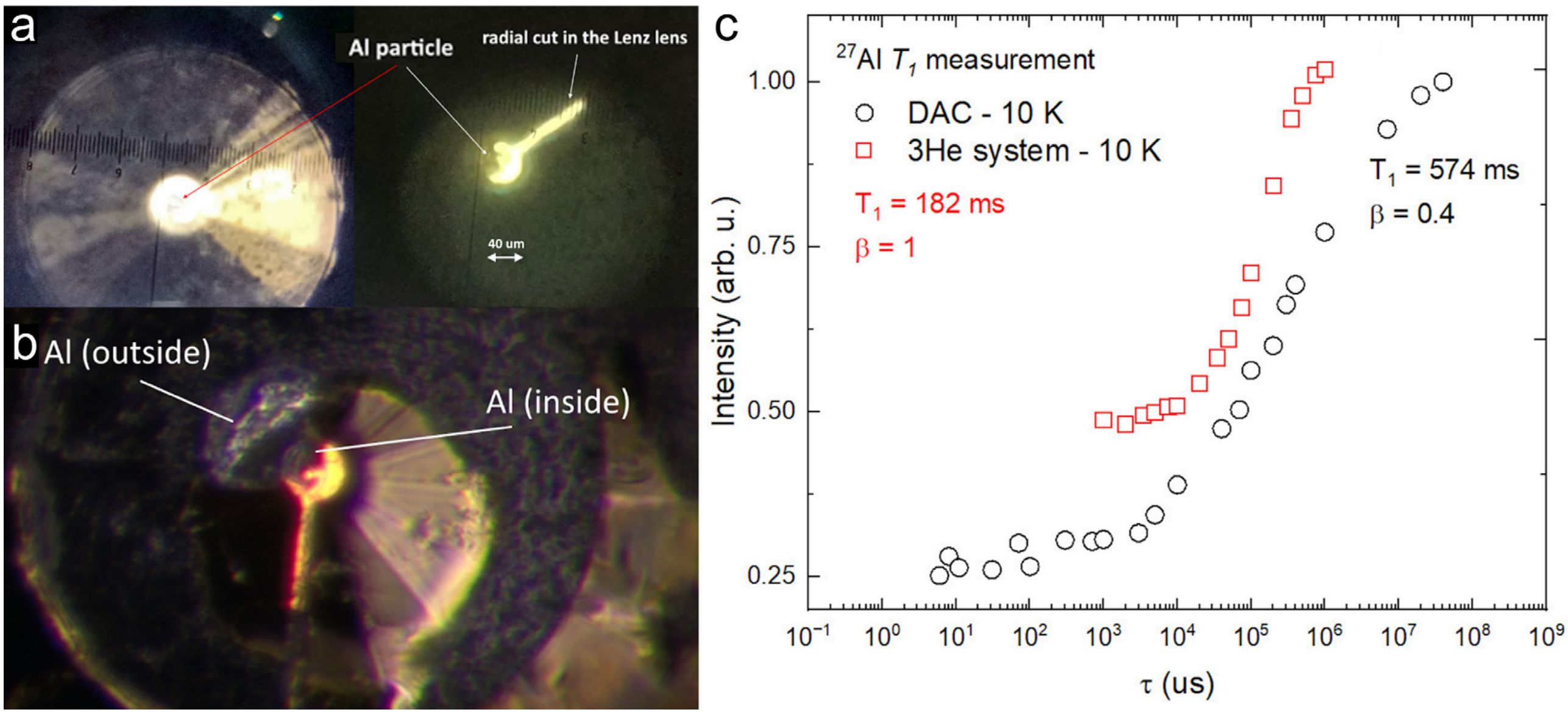

**Figure S6.** Calibration $^{27}Al$ NMR spin-echo data using a test DAC loaded with ammonia borane (AB or $NH_3BH_3$), and an aluminum ($^{27}Al$) foil, equipped with two sputtered 0.5 µm thick gold Lenz lenses. (a) Pictures of the DACs culet in reflected and transmitted light with the space filled with AB and aluminum foil. (b) Picture of the opposite anvil of the test DAC, with the Al foil in and outside the gasket. (c) Integrated intensity of the $^{27}Al$ NMR signal as function of the spin-echo delay time (τ) at 10 K.

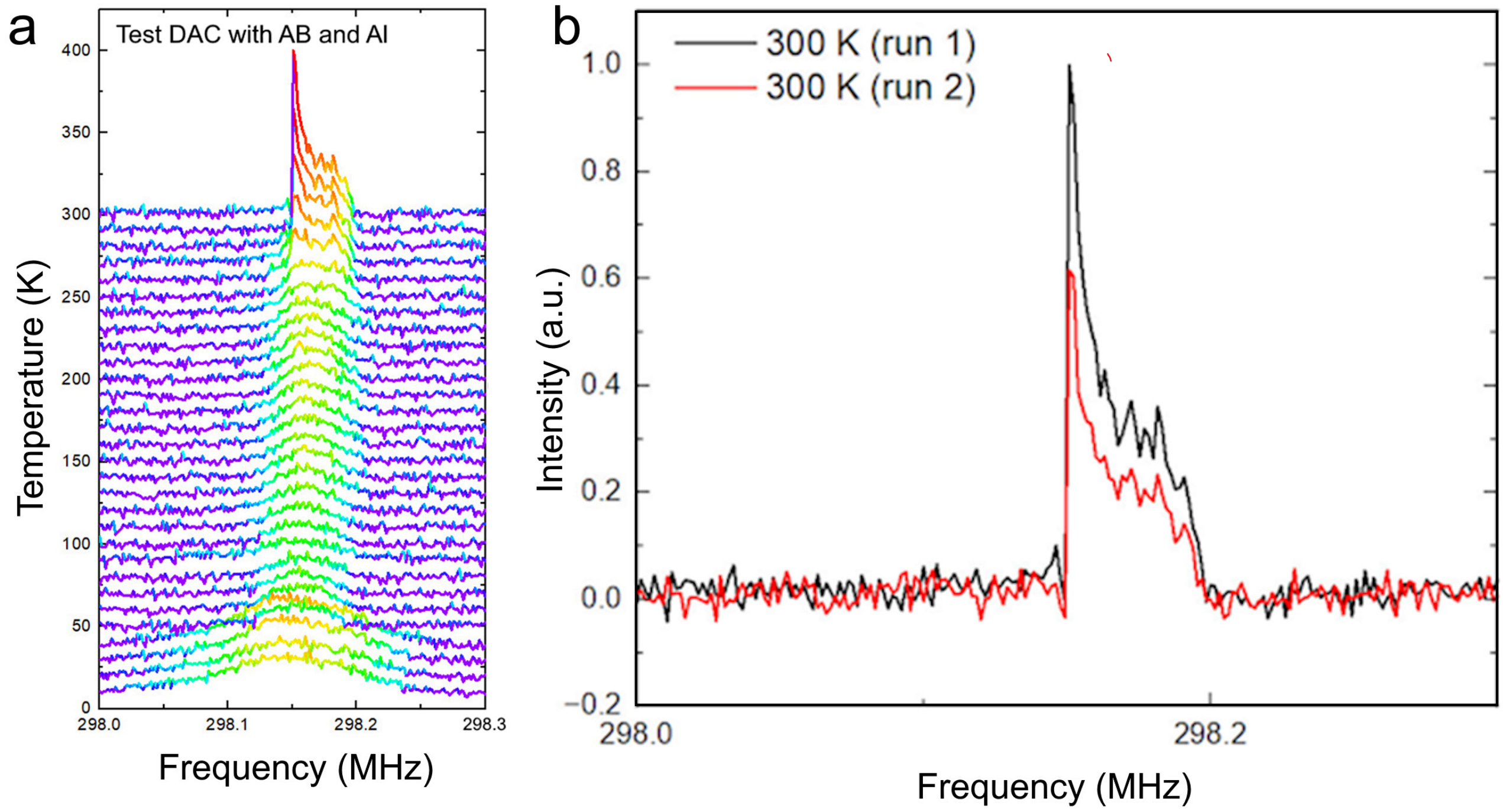

**Figure S7.** Calibration [1]H NMR data using a test DAC and loaded with ammonia borane (AB or $NH_3BH_3$), and an aluminum ($^{27}Al$) foil, equipped with two sputtered 0.5 µm thick gold Lenz lenses. (a) Stack of $^1$H NMR spectra at different temperatures during cooling from 300 K to 10 K. The signal is significantly stronger than in the case of an empty DAC. (b) $^1$H NMR spectra at 300 K before (run 1) and after cooling to 10 K and heating back to 300 K (run 2). The signal intensity is only slightly reduced due to the removal of volatile H-containing compounds from the DAC.

Figure S7 shows the $^1$H NMR measurements of the DAC, loaded with $NH_3BH_3$ and a piece of aluminum foil. The temperature dependence of the spectra is similar to the spectra measured in a conventional setup, shown in Figure S9. In contrast to the spectra of the empty DAC (Figure S8), the signal does not vanish suddenly below 230 K (Figure S7a). Importantly, after the initial cooldown to 10 K, the NMR line is almost unchanged, except for a very small loss in intensity, which we attribute to the removal of volatile $^1$H in the sample space.

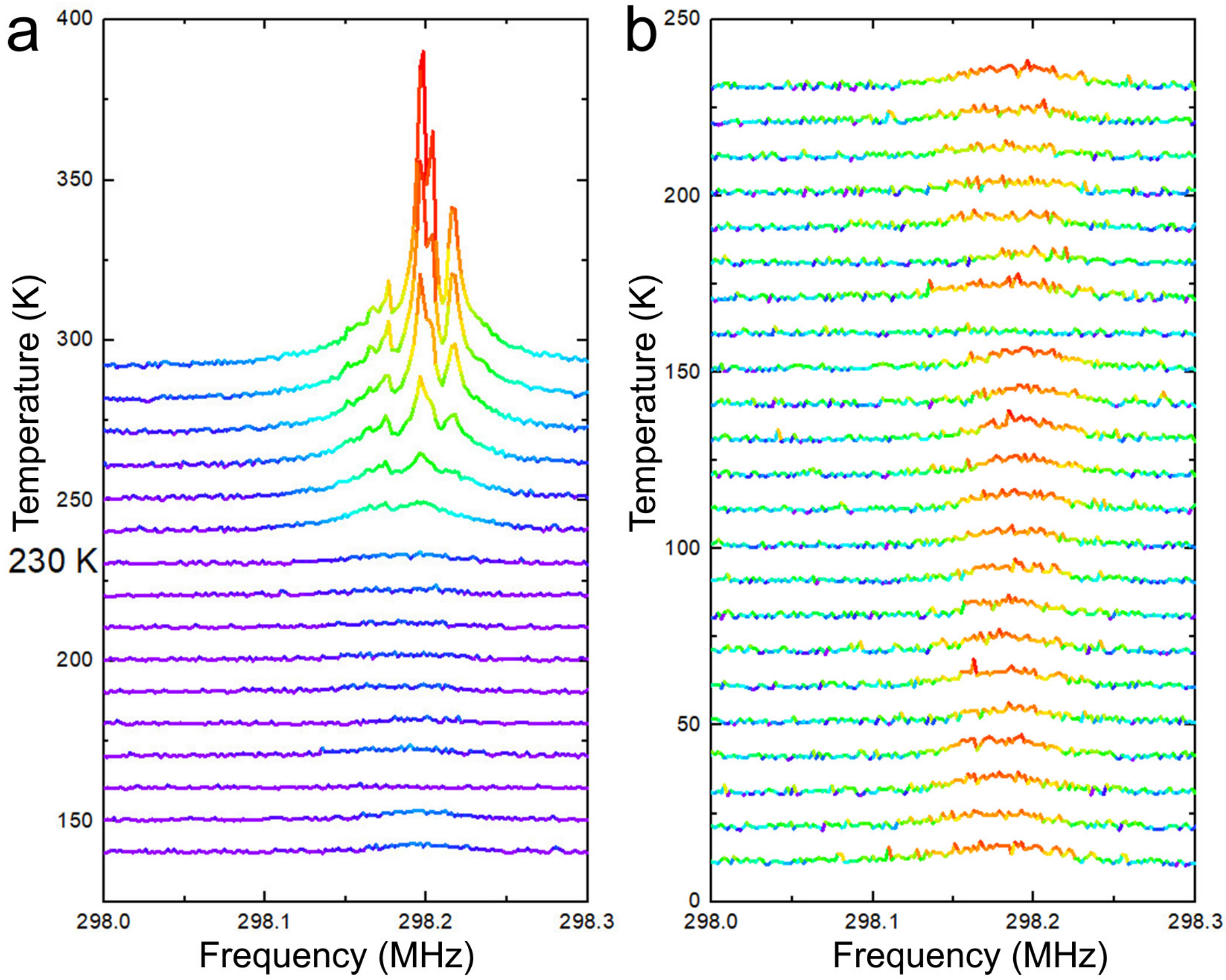

**Figure S8.** Calibration [1]H NMR data using an empty DAC equipped with two sputtered 0.5 µm thick gold Lenz lenses. The single-turn excitation coils for this DAC were printed on a ~150 µm thick teflon PCB. (a) Stack of [1]H NMR spectra at different temperatures during cooling from 300 K to 140 K, and (b) from 230 K to 10 K.

Figure S8 shows [1]H NMR measurements of an empty DAC. At temperatures above 230 K, we observe a strong [1]H signal, which quickly disappears below this temperature. In the low-temperature regime, the signal is very weak and does not show a temperature dependence. Moreover, we observed that the signal intensity did not increase again after we heated the sample back to 300 K. We interpret this behavior as contributions of volatile Hydrogen in the sample space, which disappears during the cooldown. We therefore changed our measurement protocol to first cool our sample below 200 K and then start the measurements. Importantly, this is the only DAC, which shows this behavior. All other DACs showed reproducible spectral intensities (compare Fig. S7b).

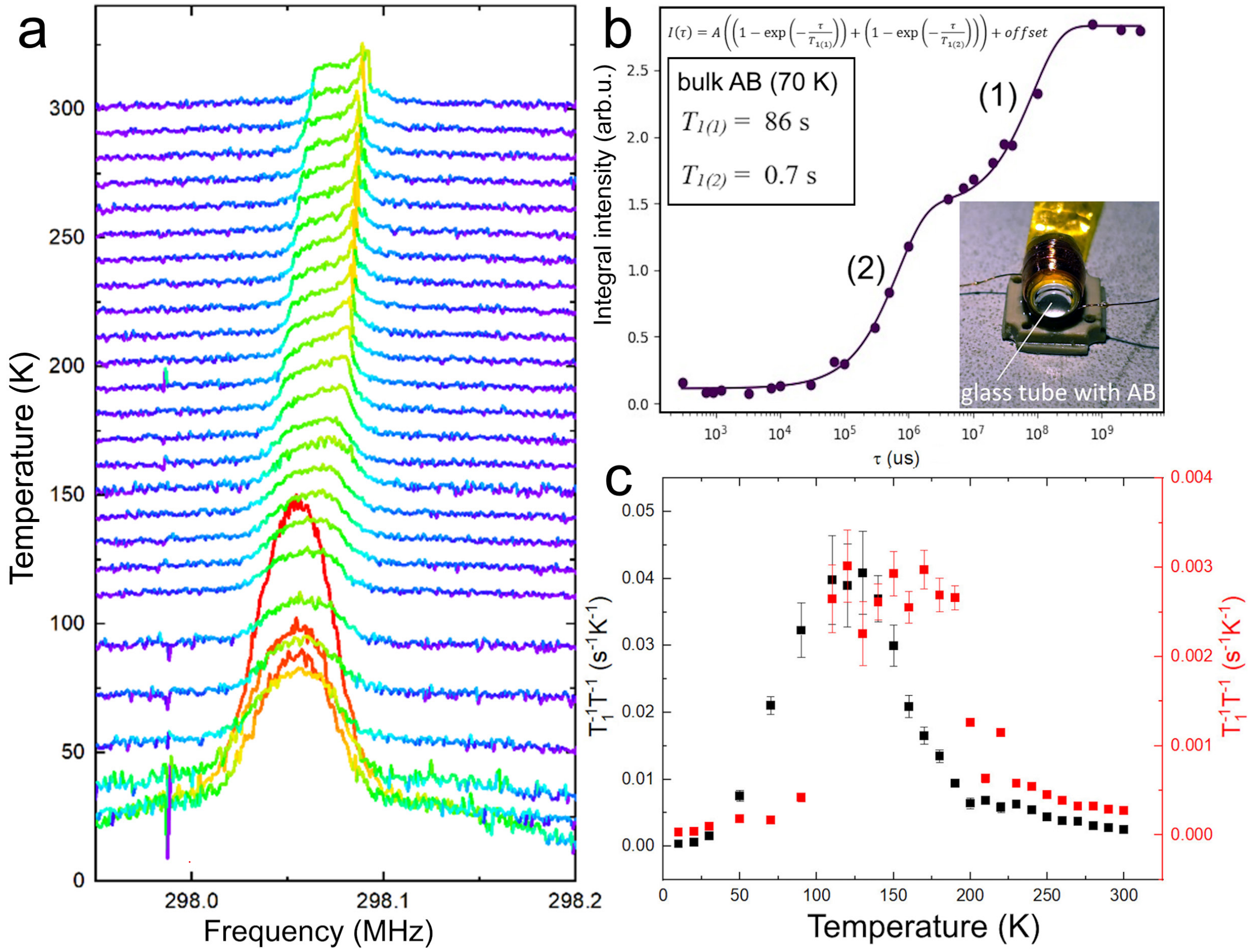

**Figure S9.** $^1H$ NMR data of a bulk $NH_3BH_3$ sample placed in a solenoid. (a) Stack of $^1H$ NMR spectra at different temperatures during cooling from 300 to 10 K. (c) Spin-lattice relaxation rate $1/TT_1$ vs $T$, obtained by the two-component fit. There is a pronounced jump in relaxation processes below 210 K in both components. (d) Integrated intensity $I$ at 70 K as a function of spin-echo delay time τ, compared to a fit describing a two-component relaxation. Inset: picture of the used solenoid with a glass ampoule (Wilmad-LabGlass, NMR sample tubes) containing AB (Sigma Aldrich).

In order to determine the temperature dependence of the $^1H$ NMR $T_1$ for $NH_3BH_3$ and contrast it to the behavior of the superhydrides, we measured a bulk sample of $NH_3BH_3$ in a conventional setup (i.e., without DAC and Lenz lenses). For this, we prepared AB powder in a glass tube and used a standard solenoid as excitation coil (inset of Figure S9b). The resulting $^1H$ NMR spectra of AB do not show a significant intensity loss at low temperatures, as is observed for the DACs. Instead, at low temperatures, the intensity increases, similarly to the behavior of the samples inside the DAC.

We measured $T_1$ by using the same procedures as described above. We note that the relaxation curves show a clear step, which indicates that the relaxation consists of two $T_1$ components. We can describe this by the following equation

$$I(\tau) = M_{z0}\left(\left(1-\exp\left(-\frac{\tau}{T_{1(1)}}\right)\right)+\left(1-\exp\left(-\frac{\tau}{T_{1(2)}}\right)\right)\right)+offset, \qquad \text{(S2)}$$

where $T_{1(1)} = 86$ s, and $T_{1(2)} = 0.7$ s at 70 K, $M_{z0}$ and $offset$ are defined as in (S1). We show the temperature dependence of both components in Figure S10c. At around 210 K, both components yield the onset of a distinct peak. Gunaydin-Sen *et al.* have observed a similar behavior in a $^{15}N$ NMR study of AB at 225 K and attribute it to a phase transition, accompanied by a change of molecular motions[6,7]. Importantly, the high-temperature regime exhibits a very different behavior from what we observe in our superhydride samples in the DACs.

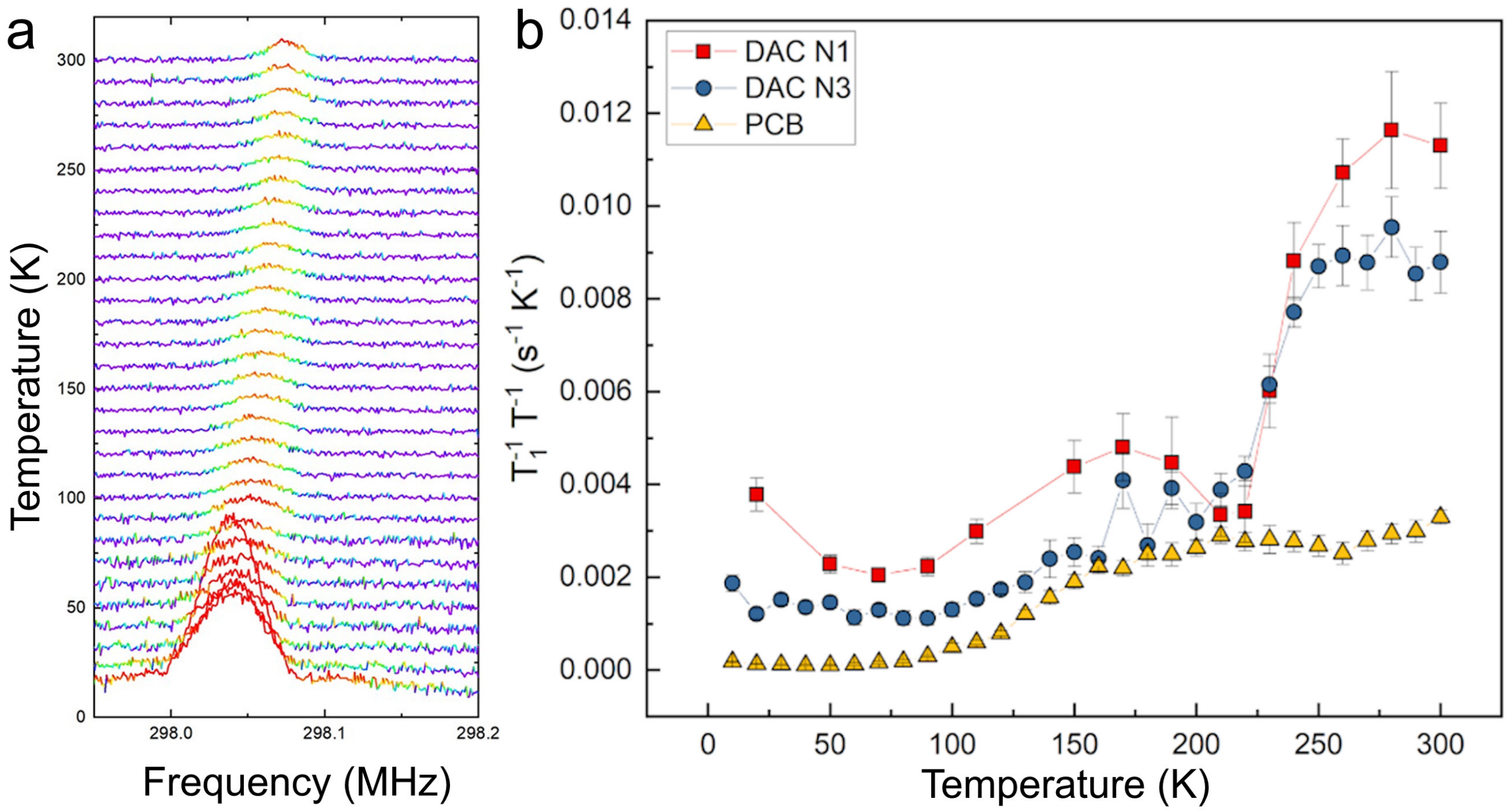

**Figure S10.** $^1$H NMR data for a piece of ~150 µm thick Teflon printed circuit board (PCB), used as a base for the single-turn excitation coil in the DAC experiments. (a) Stack of $^1$H NMR spectra at different temperatures during cooling from 300 to 10 K. (b) Dependence of the spin-lattice relaxation rate $1/TT_1$ vs $T$ of the PCB in comparison with the behavior of superhydrides in DACs N1 and N3.

To determine another contribution to the background signal, we measured the temperature dependence of the $^1$H NMR spectra of the PCB, used in the DAC preparation. We observe a shift of the spectrum towards low temperatures, which might yield a minor contribution to the overall spectral shift observed in our other measurements (Figure S10). The intensity increases strongly at low temperatures, which is a common feature in all our $^1$H NMR measurements, even in the measurements of AB in a conventional setup. Thus, the origin of this behavior may relate to residual Hydrogen in the setup. In any case, this behavior is not related to our observations for the superhydride samples.

## *5. COMSOL modeling*

The dissipated power density $P_d$ or, in other words, the surface loss density (SLD) in W/m$^2$ can be calculated in COMSOL Multiphysics 6.0[8] using

$$P_d = \frac{1}{2}(\bar{J_s} \cdot \bar{E^*}) = \frac{1}{2} t\sigma(\bar{E} \cdot \bar{E^*}) = \frac{1}{2} t\sigma|\bar{E}|^2 = \frac{1}{2\sigma} t|\bar{J}|^2 , \qquad \text{(S2)}$$

where $J_s = J{\cdot}t$ is the induced surface current density, $t$ the thickness of the metal layer, * denotes the complex conjugate, σ is the conductivity, and $\bar{E}$ is the electric-field vector. In the COMSOL simulations, we studied how the SLD and with it the electric and magnetic fields of a radio-frequency current are distributed in a thin metal layer Lenz lens deposited on a diamond anvil. In our qualitative numerical experiment, we set the thin copper layer to a thickness of 5 or 10 μm.

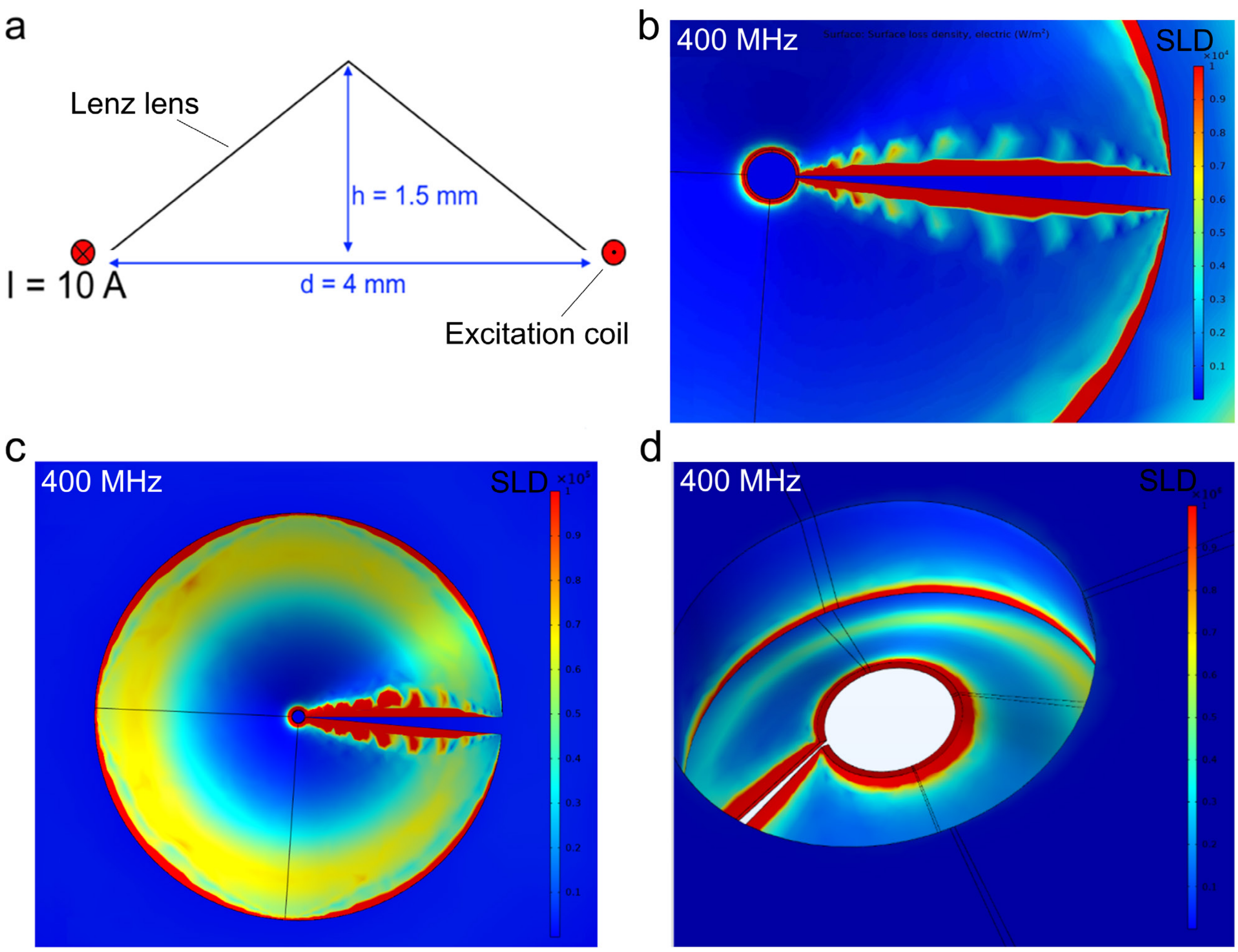

**Figure S11.** Numerical modeling of the surface-loss density distribution in a conical Lenz lens. (a) Geometric model of the conical lens used for the numerical simulations. We used a point ring current (radius is 2.5 mm) source with a current of 10 A and frequency of 400 MHz. The diameter of the hole at the top of the conical Lenz lens is 200 μm, and the thickness of the metal layer of the Lenz lens is 5 μm. (b) Current distribution (SLD) near the longitudinal slot. Current and SLD reach high values in the vicinity of the diamond anvil culet. (c) The case of a vertical shift of the ring-inductor by 0.7 mm upward. The yellow stripe corresponds to the induction on the side surface of the lens. (d) Induction in a metal gasket (300 μm thick) placed on top of a diamond anvil with a Lenz lens.

As a result of our numerical modeling, we found that a lens without a longitudinal section (slit) has a maximum SLD near the inductor and a minimum near the diamond-anvil culet. The induced current in the lens tends to suppress the electromagnetic field created by the inductor. SLD around the culet changes dramatically in the presence of a longitudinal slot (Figure 11b). Near the diamond anvil culet, the electromagnetic field now reaches its maximum. Moreover, a decrease in the diameter of the culet (that is, the cut diameter of the tip of the lens cone) causes a strong increase in the SLD and lens efficiency. Reducing the diameter by two times leads to an increase in SLD also by about two times as well.

As the frequency decreases, the current distribution becomes less localized, but the SLD maximum is still near the diamond anvil culet. A study of the influence of the inductor position shows that increasing the diameter of the inductor greatly reduces the electromagnetic field in the culet area. Downward movement also has a negative effect, but upward shift along the lens is not as critical (Figure S11c). When adding a metal seat, we observe the occurrence of electromagnetic induction in the seat material as well (Figure S11d). This phenomenon has a negative impact on the amount of energy transferred from the inductor to the Lenz lens.

## *6. X-ray diffraction data*

**Table S3.** Experimental unit-cell parameters of the lanthanum polyhydrides *P*6-$LaH_{12}$ and *C*2/*m*-$LaH_{12+x}$ used to interpret our X-ray diffraction data of the sample in DAC N1 at 165 GPa (VASP format [9]).

| ***P*6-$LaH_{12}$ (165 GPa)** | ***C*2/*m*-$LaH_{12\pm x}$ (165 GPa, x ~ 1)** |
|---|---|
| 1.0 | 1.0 |
| 4.80499 0.0 0.0 | 3.43899 0.0 0.00000 |
| -2.40249 4.16125 0.0 | 0.0 5.874 0.0 |
| 0.0 0.0 3.77270 | -1.55725 0.0 4.14299 |
| La | La |
| 2 | 2 |
| Direct | Direct |
| 0.33333 0.66666 0.75 | 0.0 0.5 0.5 |
| 0.66666 0.33333 0.25 | 0.5 0.0 0.5 |

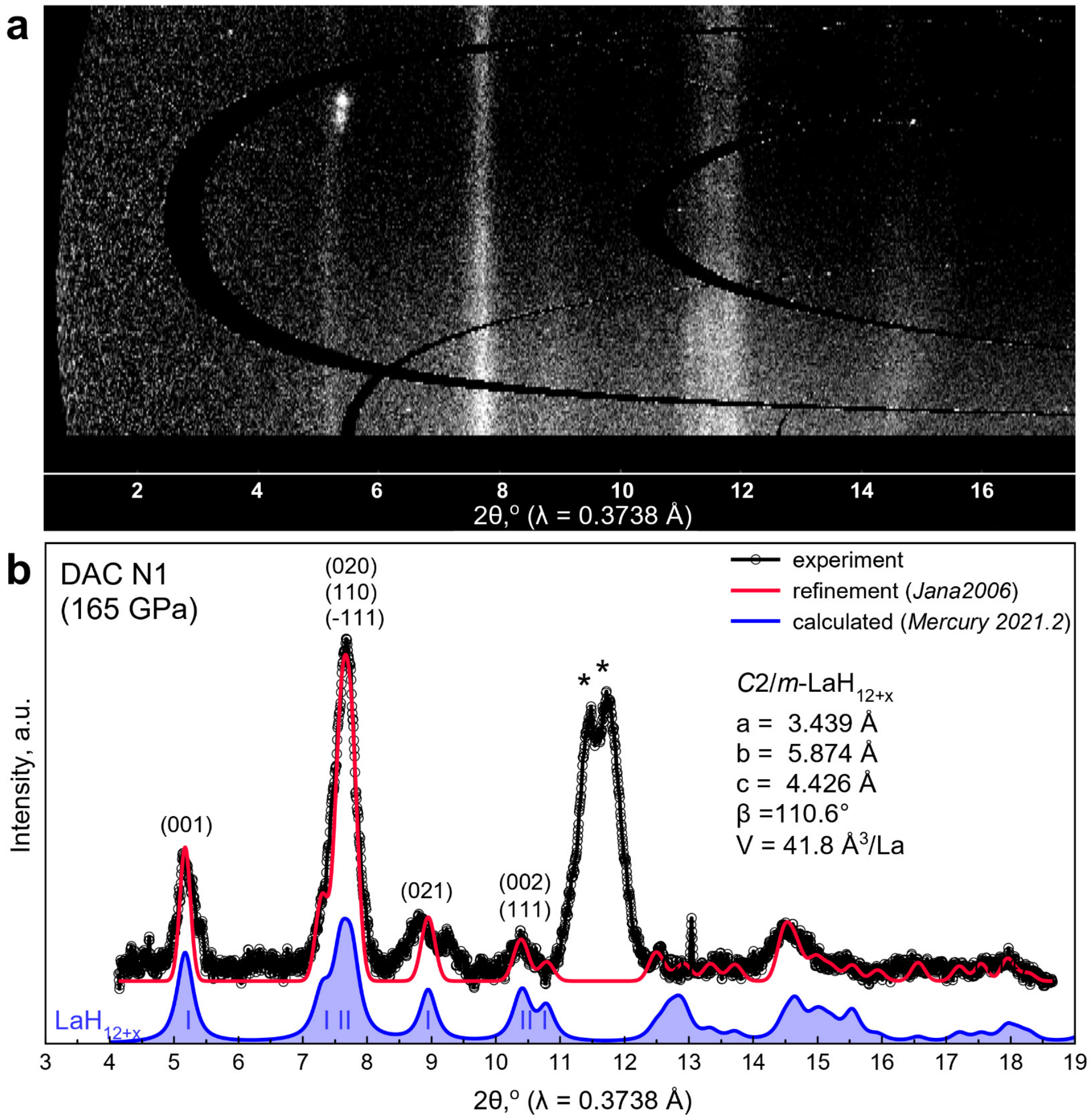

**Figure S12.** X-ray-diffraction data of the sample in DAC N1 at 165 GPa. (a) X-ray diffraction image (so called "cake") of the sample in DAC N1. Plot (b) represents the integrated intensity of the plot (a) at the corresponding diffraction angle. (b) X-ray diffraction pattern of the sample in DAC N1 and Le Bail refinement of *C2/m*-$LaH_{12\pm x}$ (x ~ 1) unit-cell parameters. Black circles are the experimental data, the red line is the Le Bail refinement, and the blue line is the XRD pattern calculated using the Mercury 2021.2 software[10]. Asterisks indicate unidentified peaks. For the refinement, we used well-known prototype structure, *C*2/*m*-$LaH_{10}$ [11], found by evolutionary structure search utilizing the USPEX code[12-15]. This structure does not explain the strong and broad reflections at 11-12 deg. Its refined unit-cell volume corresponds to a higher hydrogen content of about 12±1 H atoms for each La atom. Given the presence of the not interpreted reflections, we may speculate on a hexagonal structure of *P*6-$LaH_{12}$ (Figure 2a of the main text).

If we assume that the diffraction peak at 5.2° (Figure S12) is associated with an impurity, a description via $Fm\bar{3}m$-$LaH_{10}$ will also fail due to the low volume of such a cubic structure, V ≈ 28.5

Å$^3$/La. Instead, structures such as *C*2/*m*-$LaH_6$, *Cmme*-$LaH_6$, and $R\bar{3}m$-$LaH_6$ can, in general, explain the observed XRD pattern. They are expected from a theoretical point of view[16] and were probably obtained experimentally[17]. However, all these structures lead to a unit cell volume that is too small. Therefore, at the moment, the best description of the X-ray data is obtained using the hexagonal *P*6-$LaH_{12}$ structure.

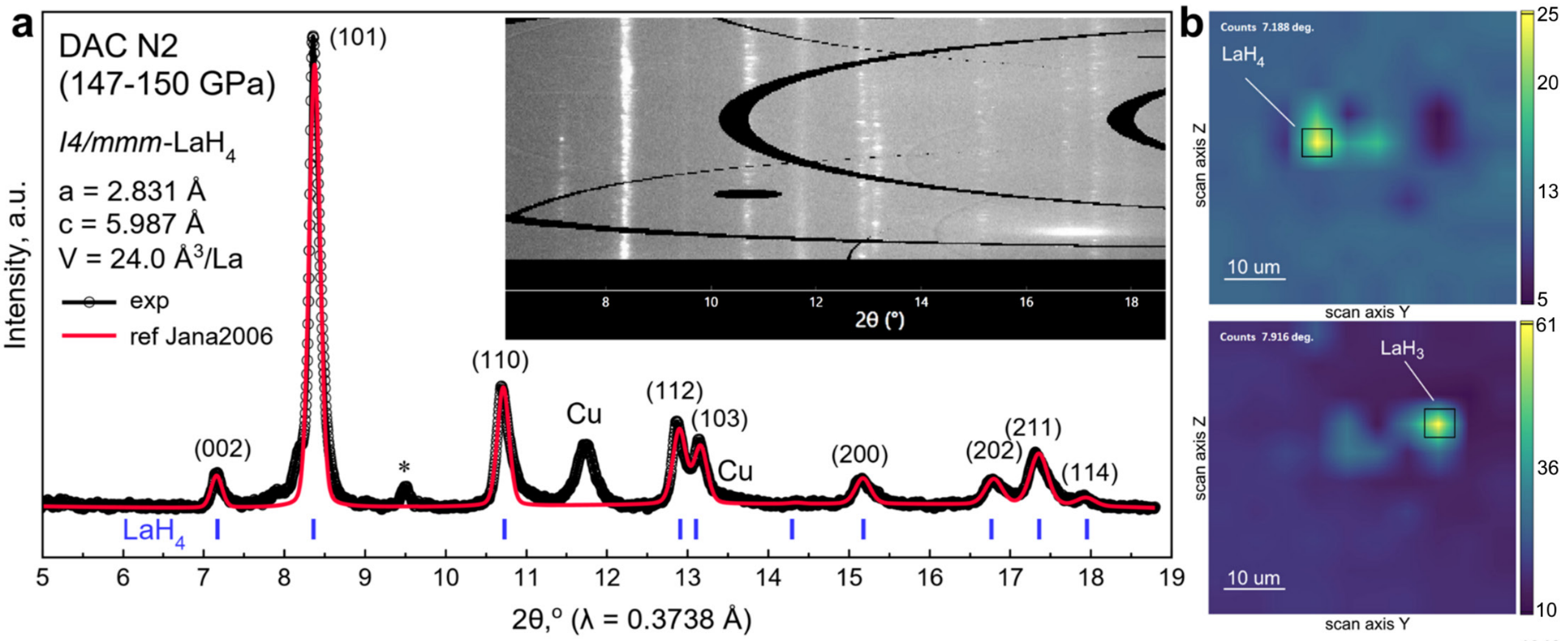

**Figure S13.** X-ray diffraction in DAC N2. (a) XRD pattern of DAC's N2 sample and Le Bail refinement of the tetragonal $LaH_4$ unit cell parameters at 147-150 GPa. Black circles are the experimental data, red line is the refinement (Jana2006 [18]), "Cu" corresponds to peaks from a copper Lenz lens. Inset: corresponding XRD image ("cake"). An asterisk (*) indicates an uninterpreted impurity signal. (b) X-ray diffraction mapping of the sample area using XDI software [19]. One can see the spatial distribution of the two main hydride phases in the sample: *I*4/*mmm*-$LaH_4$ and *Cmcm*-$LaH_3$.

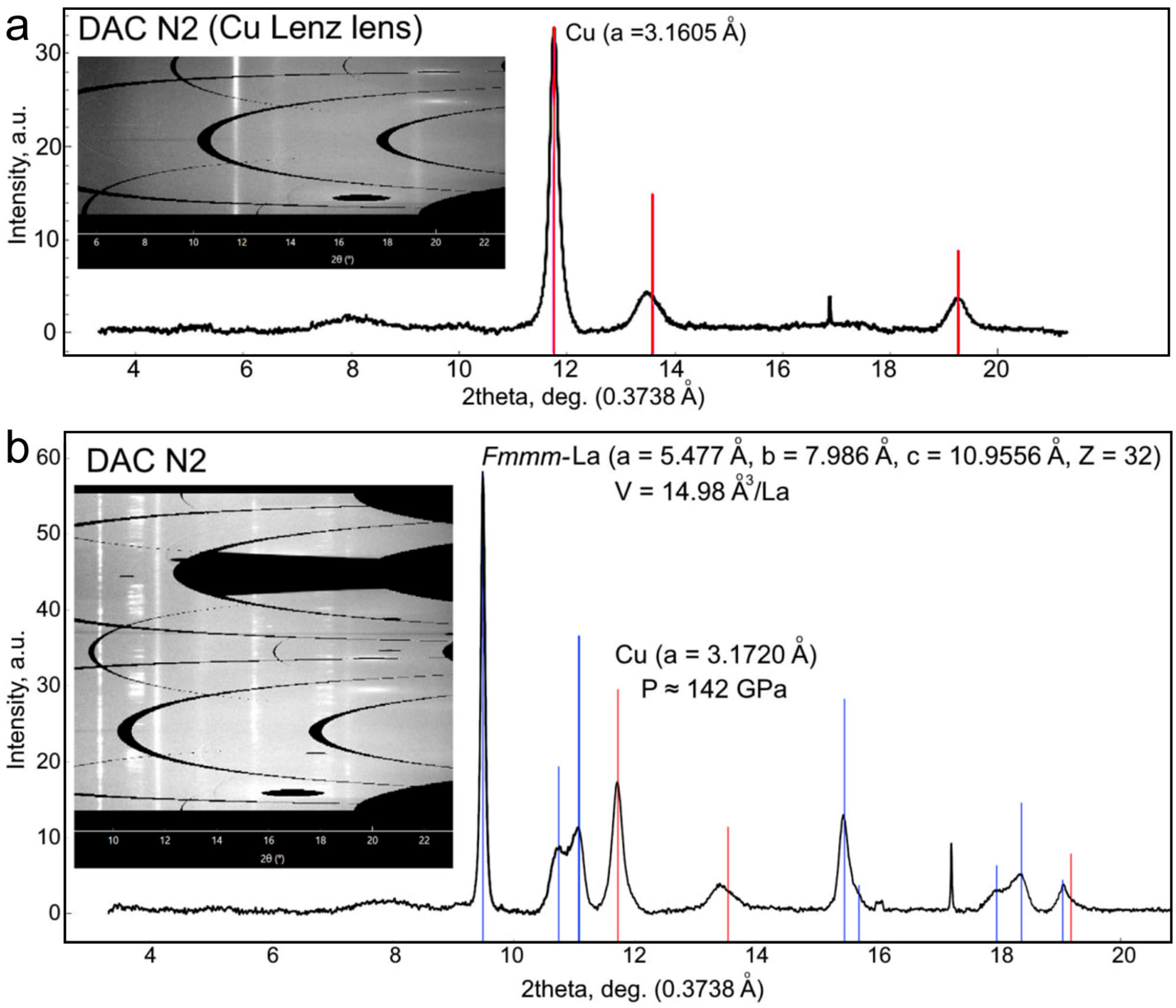

**Figure S14.** X-ray diffraction studies of DAC N2 at 147-150 GPa. This study, performed before the laser heating of the sample, shows the presence of only (a) Lenz lenses material (copper) and (b) lanthanum metal (*Fmmm*-La), the unit cell volume of which is in good agreement with earlier experiments [20].

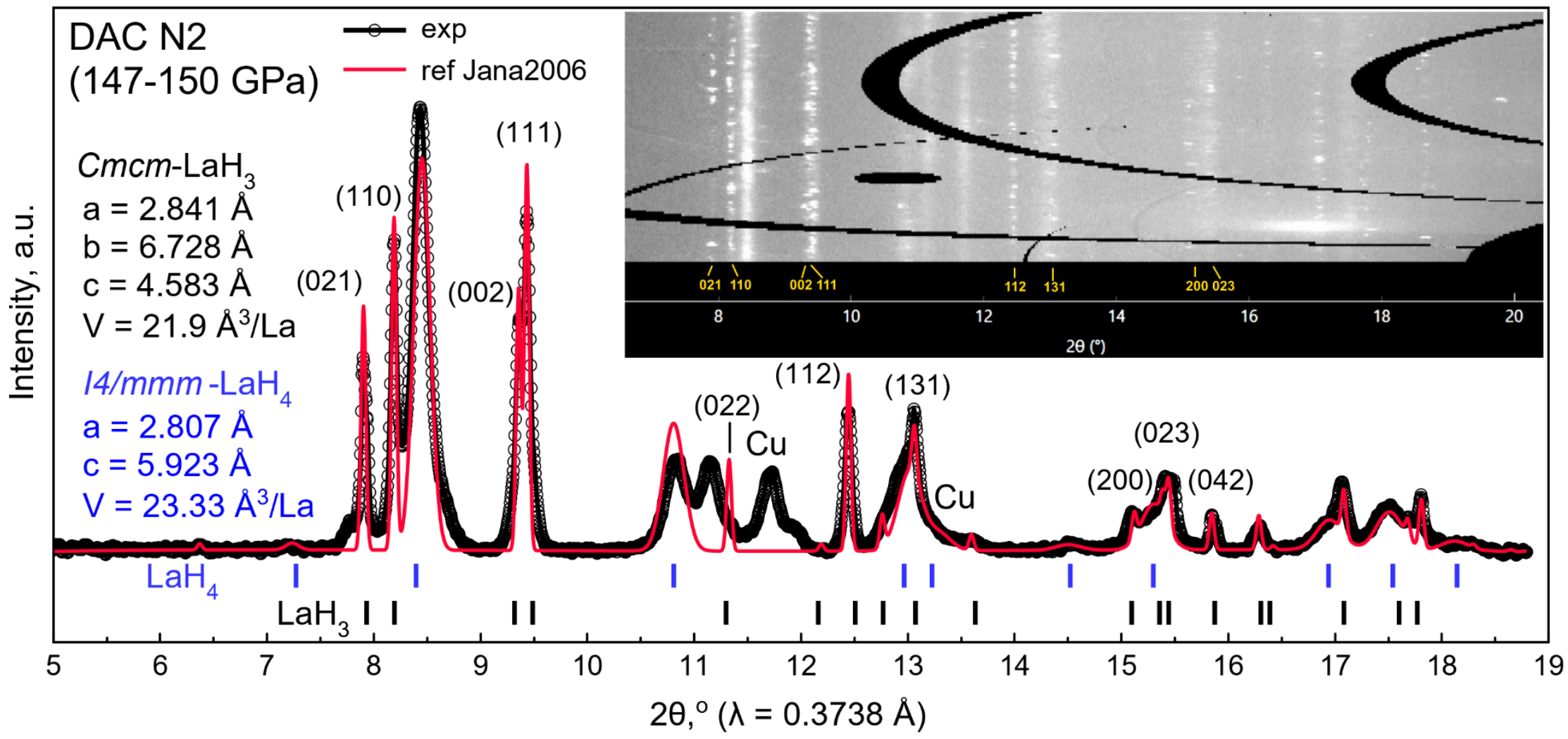

**Figure S15.** X-ray diffraction pattern of DAC N2 and Le Bail refinement of the *Cmcm*-$LaH_3$ and *I*4/*mmm*-$LaH_4$ unit cell parameters. The compounds are clearly distinguishable since the X-ray diffraction pattern of $LaH_3$ has a "spot-like" coarse-crystalline character. *Cmcm*-$LaH_3$ can be predicted theoretically using an evolutionary search for stable crystal structures. At 100 GPa the $LaH_3$ is 0.47 meV/atom above the convex hull of the La-H system at 100 GPa which is not so far [16].

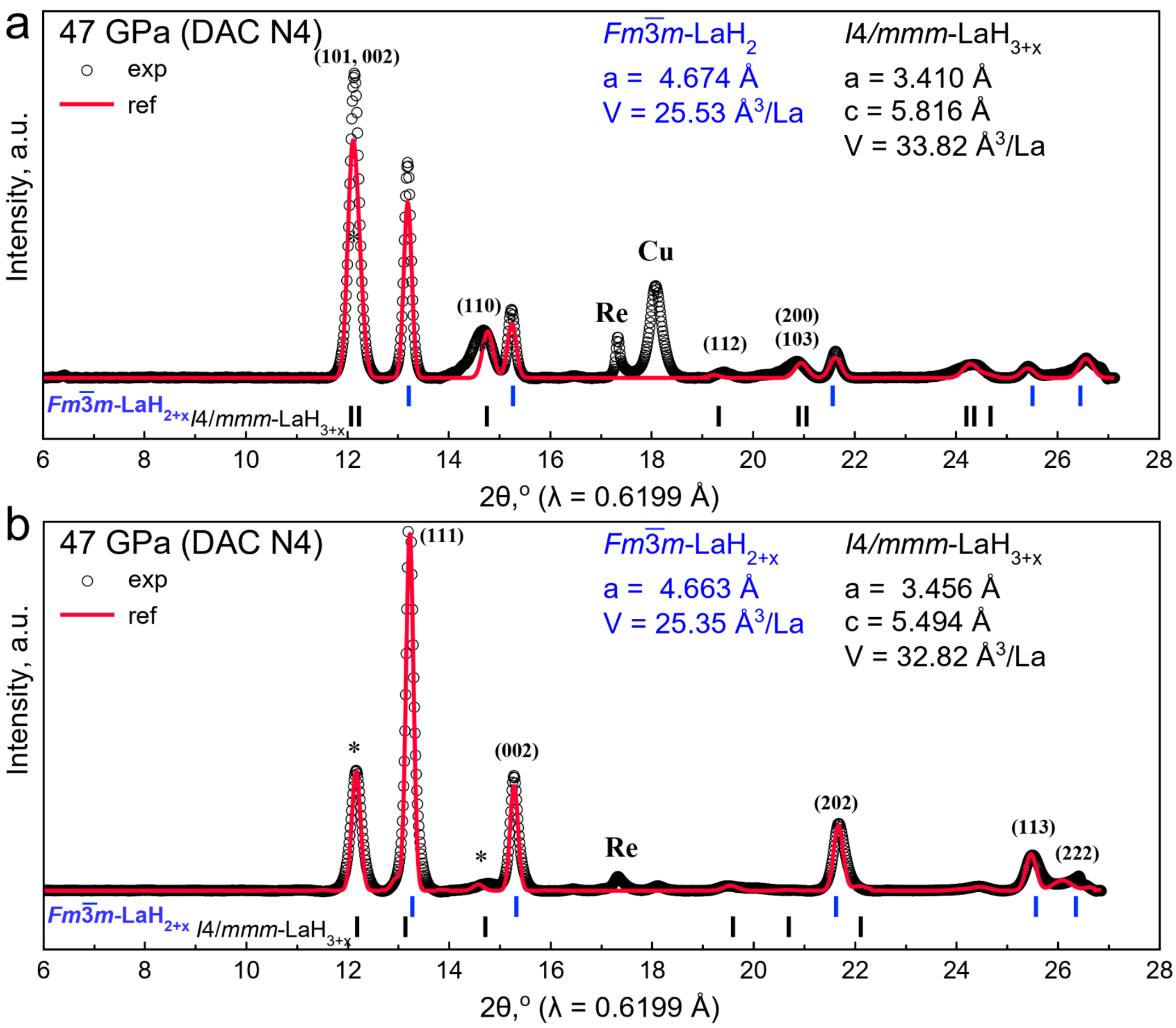

**Figure S16.** X-ray diffraction pattern of DAC N4 and Le Bail refinement of the *Fm*$\bar{3}$*m*-$LaH_{2+x}$ and *I*4/*mmm*-$LaH_{3+x}$ unit cell parameters before the $^1$H NMR experiment. Black circles are the experimental data, red line is the refinement (Jana2006 [18]), "Re" corresponds to peaks from a rhenium gasket. After low-temperature NMR measurements pressure in the DAC N4 dropped to 19 GPa. Panels (a) and (b) correspond to different regions in the sample.

*7. Raman spectroscopy*

The purpose of the Raman spectroscopy was to establish the pressure before and after NMR experiments in DACs N0-N4. We found that during sample cooling to 10 K with subsequent warming up back to the room temperature, the pressure change in the DACs is within 5-10 GPa. Examples of pressure measurements before the experiment (the Akahama scale [21]) are shown in Figure S17.

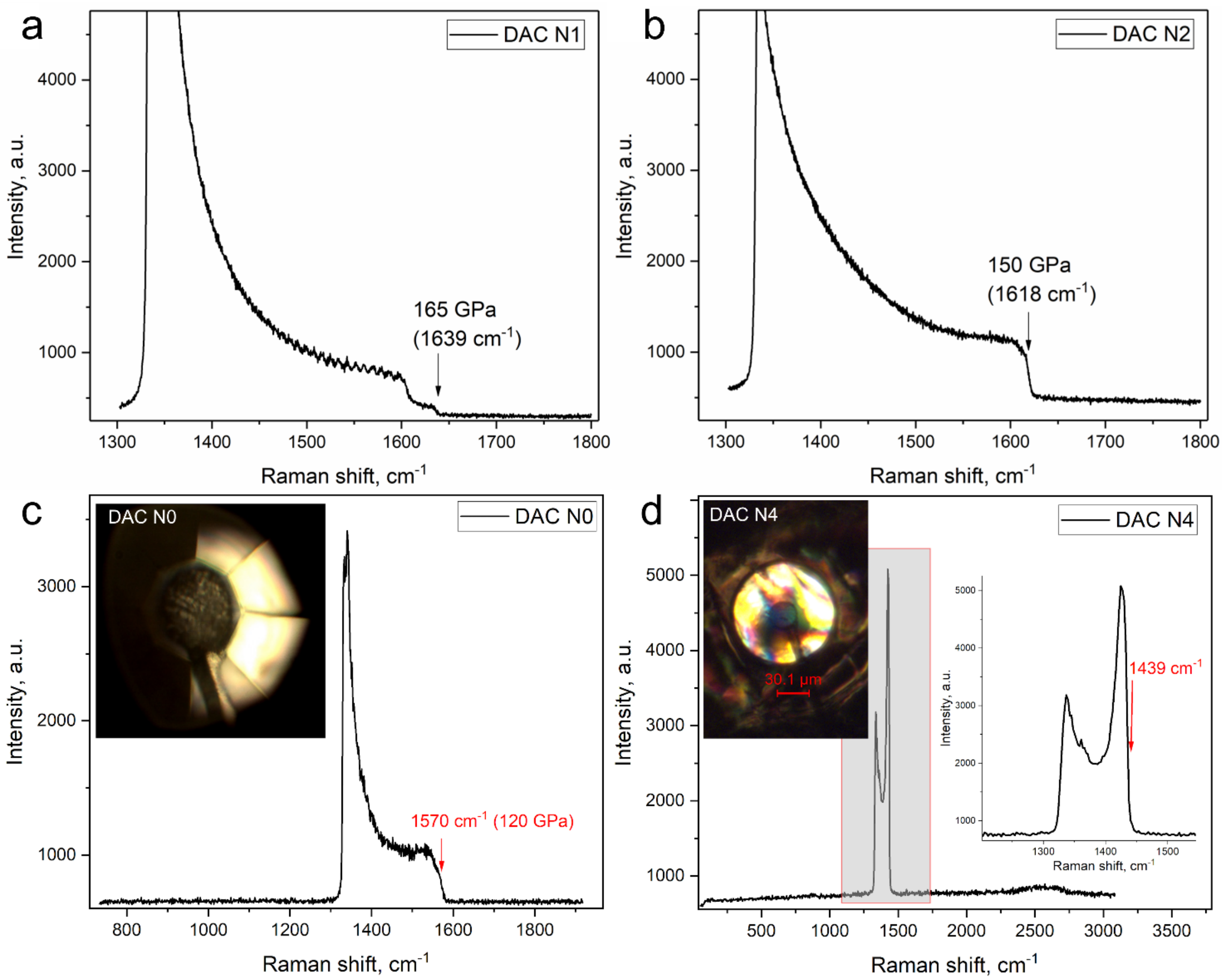

**Figure S17.** Raman spectra of the samples in (a) DAC N1 (b), DAC N2 (c), DAC N0, and (d) DAC N4 before the NMR experiment. We used lasers with wavelengths of 633 nm and 532 nm for excitation. The Akahama scale[21] was applied to determine the pressure. After the experiment at 5 K, we found that the pressure in DAC N4 dropped to 19 GPa. Insets: Photographs of the loaded high-pressure chambers of (c) DACs N0 and (d) DAC N4.

## *8. Additional NMR data*

### *$^{1}H$ NMR of $LaH_{12}$ (DAC N1)*

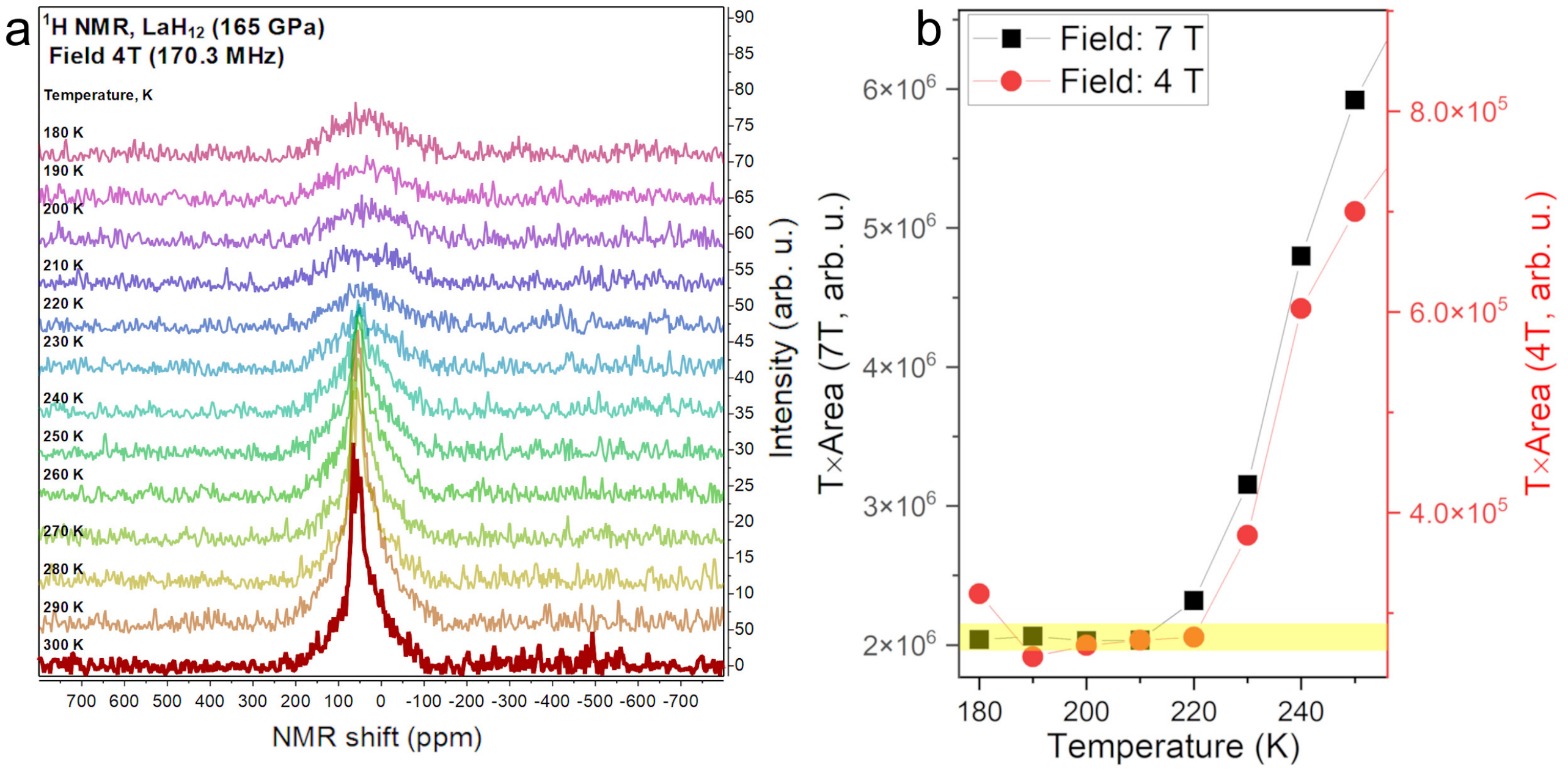

**Figure S18.** NMR characteristics of $LaH_{12}$ in DAC N1 sample in magnetic fields of 7 T and 4T at different temperatures. (a) Stack of $^{1}H$ NMR spectra at various temperatures from 300 K to 180 K. (b) Superconducting transition in $LaH_{12}$ at 165 GPa in 4 T and 7 T determined by the drop in the $^{1}H$ NMR peak area times temperature. The offset transition temperature of 220 K at 4 T is noticeably higher than at 7 T (211 K).

Figure S18 shows a comparison of the $^{1}H$ NMR data of the sample in DAC N1 between 4 and 7 T. The temperature dependence of the $^{1}H$ NMR spectra at 4 T (Figure S18a) is qualitatively similar to the temperature dependence at 7 T (compare Figure 3 in the main text). A comparison of the spectral properties at 4 and 7 T reveals a slightly higher $T_c$ at the lower field of 4 T, as expected.

### *Spin-lattice relaxation time ($T_1$) of $LaH_{12}$ (DAC N1)*

NMR is an effective method for determining the superconducting gap $\Delta(T)/k_B$, since in fully gapped superconductors at temperatures below $T_C$, the spin-lattice relaxation rate is described by the exponential law $1/T_1 \propto e^{-\frac{\Delta(0)}{k_B T}}$ [22]. A fit or our $1/T_1$ data of $LaH_{12}$ with this simple exponential function gives $\Delta(0)/k_B$ = (307 ± 41) K, which equates to (26.5 ± 3.5) meV (Figure S19a). This value of the superconducting gap is several times less than theoretical estimates of 60-80 meV[16,23]. The probable cause of this discrepancy is the multiphase contribution to the $^{1}H$ NMR signal of DAC N1:

$$\frac{1}{T_1} = \frac{1}{T_1^{SC}} + \frac{1}{T_1^{N}} = Ae^{-\frac{\Delta(0)}{k_B T}} + CT, \qquad \text{(S3)}$$

where we assumed the validity of the Heitler-Teller equation for the non-superconducting H-containing part of the system, providing the $T_1^N$ term. In this case, the fit gives $\Delta(0)/k_B$ = (427 ± 49) K, which equates to (37 ± 4) meV, and $R_\Delta = 2\Delta(0)/k_B T_c$ = (3.76 ± 0.42) (Figure S19c), in line with the weak-coupling BCS limit (3.54), but less than what the Migdal-Eliashberg theory predicts for the $LaH_{10}$ ($R_\Delta \sim 5.5$[16]). A further improvement of the accuracy of the SC gap estimate requires hydride samples of higher quality and mass. We achieve a better agreement with the prediction when neglecting the anomaly below 227 K. This anomaly may be a Hebel-Slichter peak (Figure S20). Using Eq. (S3) gives $\Delta(0)/k_B$ = (672 ± 42) K, which equates to (57.9 ± 3.6) meV, and $R_\Delta = 2\Delta(0)/k_B T_c$ = (5.16 ± 0.32), with $T_c$ = 260 K at 7 T (Figure S19d).

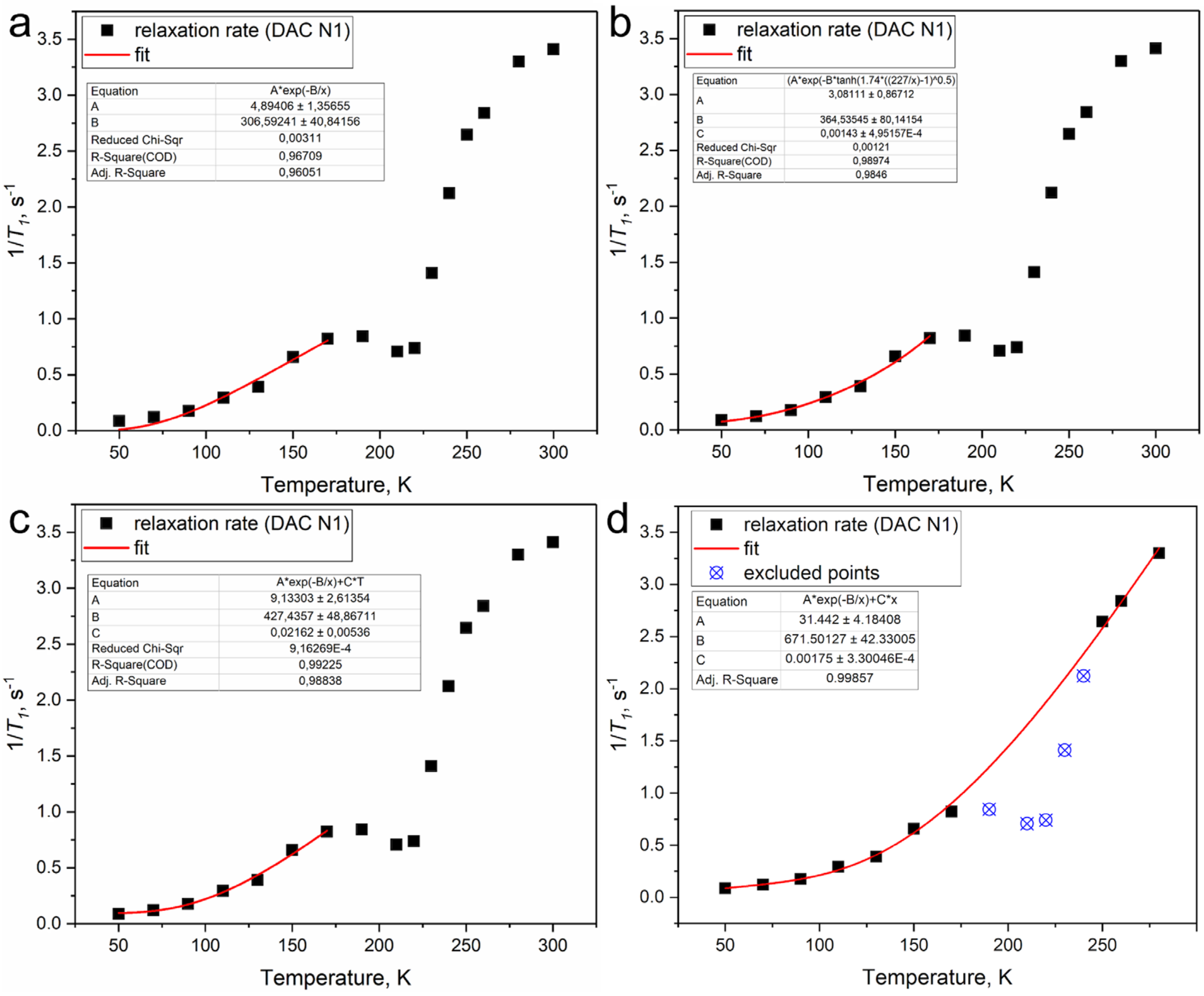

**Figure S19.** Different fits to the temperature-dependent [1]H NMR $1/T_1$ data at $T < T_C$ for the sample in DAC N1 at 165 GPa and 7 T. (a) Using a simple exponential function for the data below 170 K gives $\Delta(0)/k_B$ = (307 ± 41) K. (b) Using an interpolation formula for $\Delta(T)/k_B$ proposed by Gross et al.[24] for data below 170 K gives $\Delta(0)/k_B$ = (365 ± 80) K. (c) Considering the presence of a non-superconducting phase ($1/T_1 \propto CT$, Eq. S3) below 170 K results in $\Delta(0)/k_B$ = (427 ± 49) K. (d) Excluding data around the anomaly at 227 K and using Eq. S3 as well gives $\Delta(0)/k_B$ = (672 ± 42) K, which is the best agreement with the theoretical predictions.

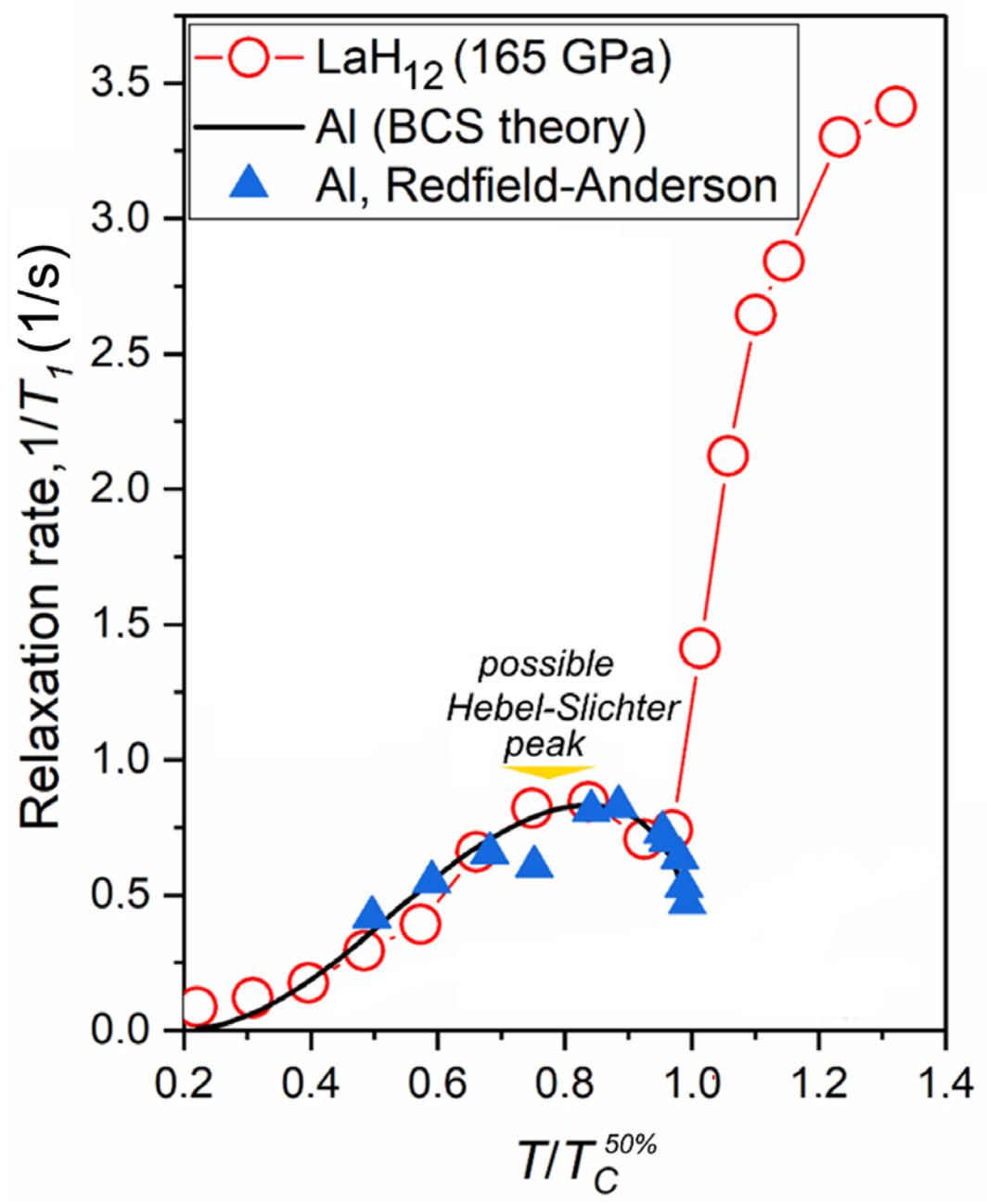

**Figure S20.** Temperature dependence of the [1]H NMR spin-lattice relaxation rate of $LaH_{12}$ (DAC N1) at 7 T (red circles) and aluminum (blue triangles, from Ref. [21]) versus reduced temperature $T/T_c^{50\%}$. The black line corresponds to the prediction of BCS (dirty limit)[25].

Below 230 K, the temperature dependence of $1/T_1$ of the sample in DAC N1 reveals a feature (Figure S20), which compares well to the results measured by Anderson in aluminum[21]. Hebel and Slichter were the first to describe this behavior of $1/T_1$ of aluminum in the superconducting state by applying the BCS theory[26,27]. If the feature observed in Figure S20 is indeed a Hebel-Slichter peak, this would confirm the gapless (likely BCS-type) nature of the superconducting order parameter in $LaH_{12}$. Although we note that, generally, the Hebel-Slichter peak is quite rarely observed experimentally and so far, this is the only sample, in which we observe this feature.

*$^1H$ NMR $LaH_{\sim3}$ and $LaH_4$ (DAC N2)*

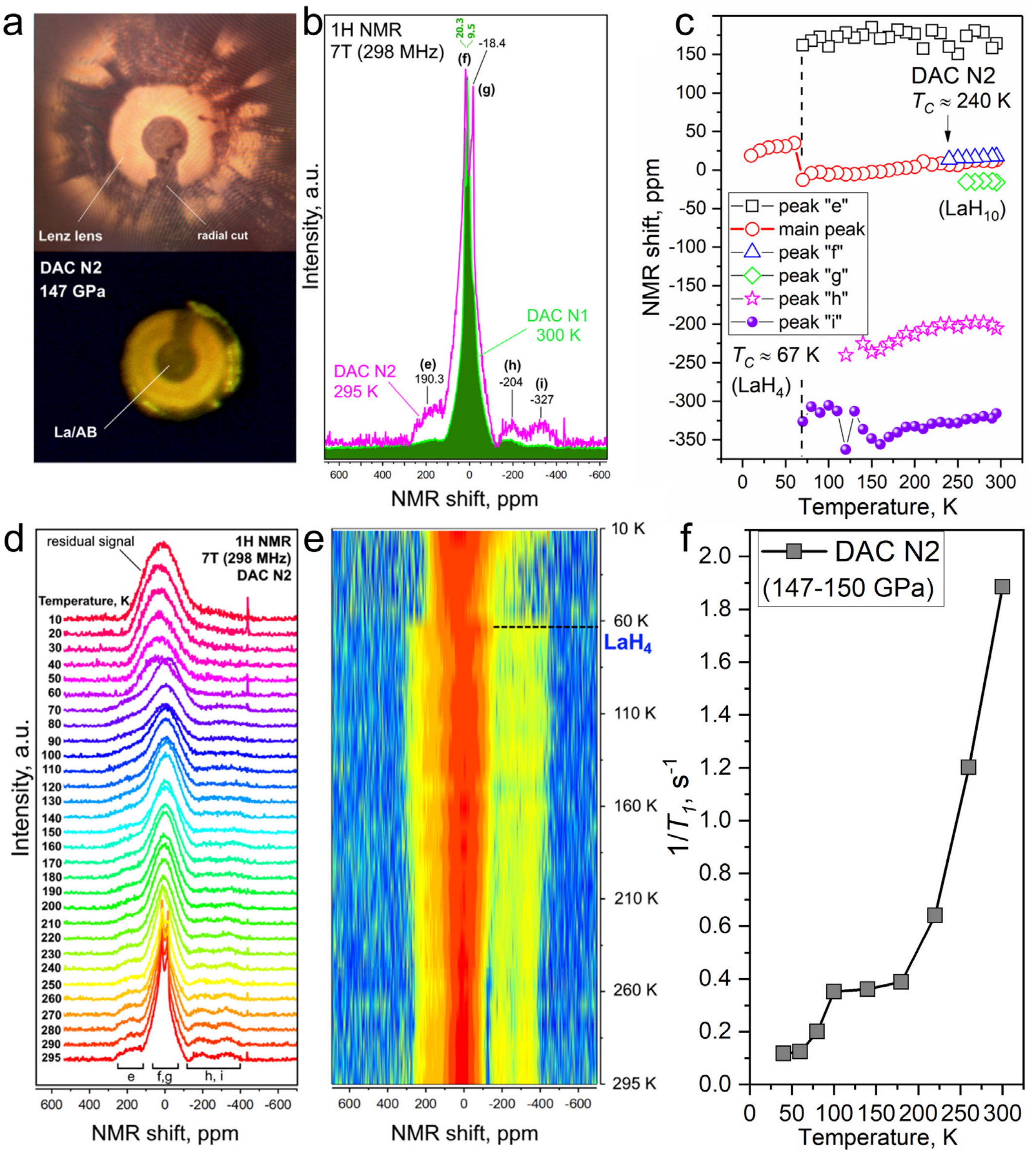

**Figure S21.** NMR data of the sample in DAC N2. (a) Pictures of DAC N2 culet at 147 GPa. The geometry of the Lenz lens is clearly visible and remains unchanged under compression. (b) Comparison of $^1H$ NMR spectra of the samples in DAC N1 (green) and DAC N2 (magenta) at a temperature of 300 and 295 K, respectively. In addition to a signal similar to that observed for the sample in DAC N1 ("f, g"), a series of broad signals "e, h, i" is observed, which may correspond to $LaH_4$, since they disappear below 67 K. (c) Temperature dependence of the NMR shift of each component of the complex NMR signal. Two peaks, "f" and "g" disappear below 240 K, while the signals "e", "h", and "i" disappear below 60-70 K. (d) Stack of $^1H$ NMR spectra of the sample in DAC N2 in 7 T at various temperatures from 295 to 10 K. (e) $^1H$ NMR intensity colormap of the data presented in panel (d). Red regions have the highest intensity and blue ones the lowest intensity. The sharp change in the signal below 60-70 K may point to the SC transition in $LaH_4$. (f) Temperature dependence of $1/T_1$ of the sample in DAC N2. Below about 100 K, $1/T_1$ drops. The midpoint of the transition is about 65-75 K.

Despite the small sample size, the NMR signal of the sample in DAC N2 has acceptable quality (Figure S21). The disappearance of two narrow $^1$H NMR peaks "f" and "g" at temperatures below 250 K may be caused by the presence of small fractions of higher lanthanum hydrides ($LaH_{10}$ or $LaH_{12}$), missed due to the large scanning step (5 μm) used in the X-ray diffraction experiment. Around 60-70 K, all NMR signal characteristics (frequency shift, height, width, and area) undergo a pronounced change. This may be associated with the superconducting transition of $LaH_4$ (Figure S21c,e,f).

The broad and weak NMR signals "e" and "i" (Figure S21c) do not yield significant features in their temperature dependences, and disappear between 60 and 70 K. The same holds for the even weaker signal "h", which disappears below 110-120 K (Figure S21c). The relaxation time $T_1$ of the sample in DAC N2 is about 530 ms, and $TT_1 \approx 160$ sK at 300 K. The relaxation rate yields a drop in the range of 60-70 K, which, however, starts much earlier, at around 100 K (Figure S21f).

We studied the spin-spin relaxation time $T_2$ of the sample in DAC N2 in a range from 5.5 K to 270 K. We found that $T_2$ is 32 – 47 μs and almost independent of temperature (Supporting Table S4). This situation of $T_2 << T_1$ is typical for solid-state NMR of metal hydrides (e.g., Ref. [28]).

**Table S4.** Spin-spin relaxation time $T_2$ of the sample in DAC N2, (147-150 GPa) in 7 T at different temperatures.

| Temperature, K | $T_2$, μs | *Error in $T_2$, ε(%)* |
|---|---|---|
| 270 | 47.1 | 32.5 |
| 200 | 39.55 | 42 |
| 100 | 32.05 | 19 |
| 5.5 | 37.61 | 41.4 |

### *$^1$H NMR of $LaH_{\sim 3}$ (DAC N4)*

$^1$H NMR studies of lower lanthanum hydrides were carried out using DAC N4 loaded with $LaH_{3-x}$ and ammonium borane in 8 T magnetic field. The DAC was originally intended for research at high pressures, had a culet of 75 μm, and was heated using an IR laser at 150 GPa. However, a failure of one of the anvils led to a pressure drop to 47 GPa. After the low-temperature NMR experiment the pressure decreased further to 19 GPa. According to X-ray diffraction data, the sample consists of two phases: cubic (*fcc*) $LaH_{2+x}$ and, probably, tetragonal (*I4/mmm*) $LaH_{3+x}$ (Figure S16). Both phases are not high-$T_c$ superconductors. To simplify the further discussion, and considering the starting compound for the synthesis, we will henceforth refer to the sample in DAC N4 as $LaH_3$.

The $^1$H NMR examination of this sample evidenced its non-superconducting nature (Figure S22). During the cooling, the NMR signal broadens significantly and experiences a slight frequency shift of about 26 ppm. This emphasizes the absence of a pronounced Knight shift in many lanthanum hydrides. The width (FWHM) of the NMR signal below 100 K reaches 40-50 kHz, which corresponds to the width of the residual signal in the DACs N1-N3 at low temperatures. The spin-lattice relaxation time of $LaH_3$ in DAC N4, measured at 5 K, is $T_1 \approx 100$ s, and $TT_1 \approx 500$ sK. Comparing with the low-temperature behavior of $T_1$ in DACs N1 and N2, the differences in $TT_1$ are small: we found $TT_1$ = 573 sK at 50 K in DAC N1, and 340 sK at 40 K in DAC N2. This is not surprising, given the fact that SC phases make virtually no contribution to the relaxation rate at much below $T_C$. Therefore, the spin-lattice relaxation at low temperatures is mainly due to the residual non-SC $LaH_3$ phase and other H-containing phases.

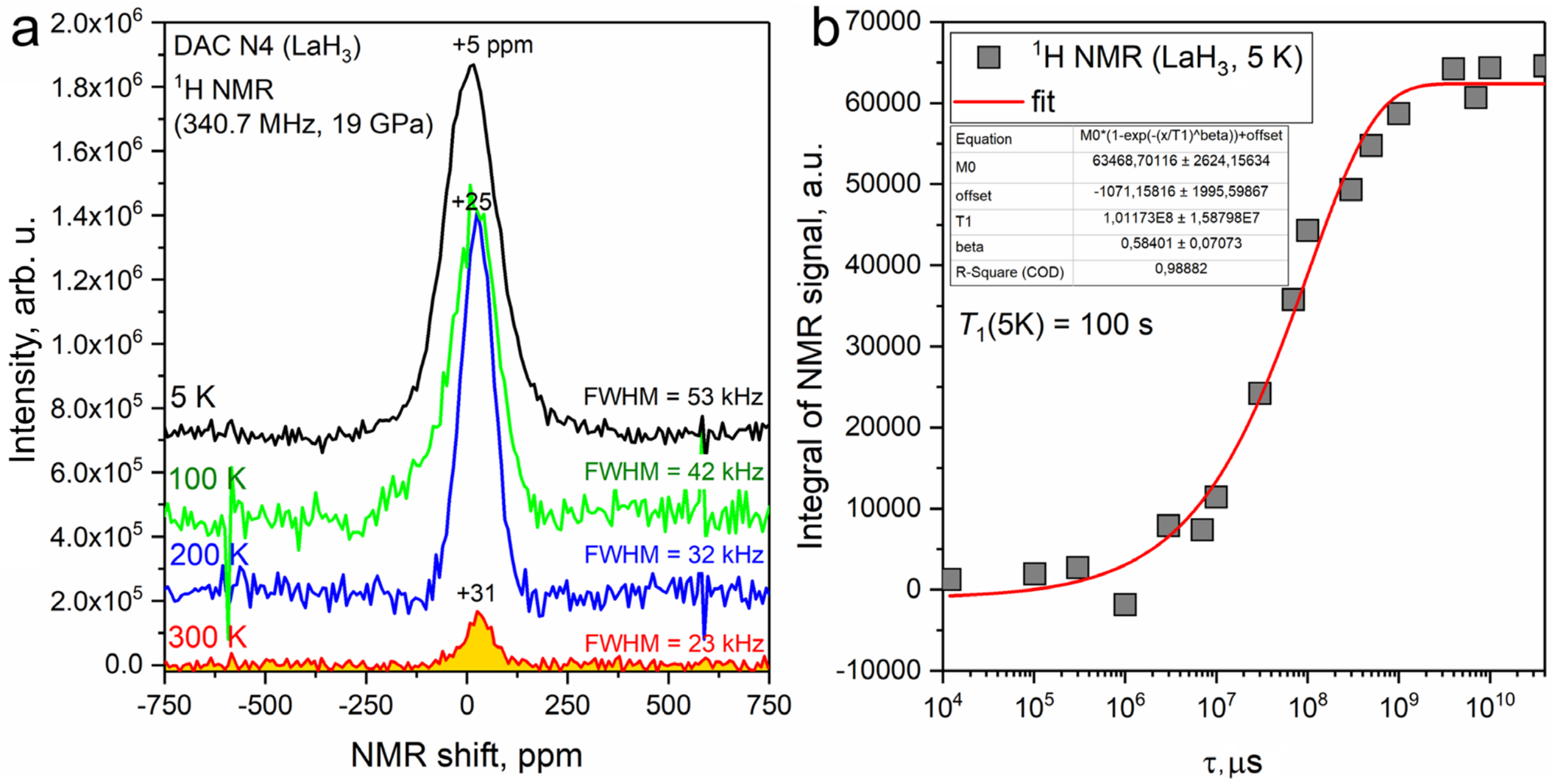

**Figure S22.** $^1H$ NMR of the sample in DAC N4. (a) Stack of $^1H$ NMR spectra at 19 GPa in an 8 T field at 5, 100, 200, and 300 K. (b) Dependence of the integral intensity of the $^1H$ NMR signal on the interpulse delay time τ within the spin-lattice relaxation experiment at 5 K. The found relaxation time $T_1$ is about 100 s.